\documentclass[aps,prb,twocolumn,superscriptaddress,showpacs]{revtex4}

\usepackage{graphicx}
\usepackage{afterpage}

\begin{document}


\title{Kondo effect and absence of quantum interference effects in the
charge transport of cobalt doped iron pyrite}


\author{S. Guo}
\affiliation{Department of Physics and Astronomy, Louisiana State
University, Baton Rouge, Louisiana 70803 USA}

\author{D.P. Young}
\affiliation{Department of Physics and Astronomy, Louisiana State
University, Baton Rouge, Louisiana 70803 USA}

\author{R.T. Macaluso}
\affiliation{Department of Chemistry, Louisiana State
University, Baton Rouge, Louisiana 70803 USA}

\author{D.A. Browne}
\affiliation{Department of Physics and Astronomy, Louisiana State
University, Baton Rouge, Louisiana 70803 USA}

\author{N.L. Henderson}
\affiliation{Department of Physics and Astronomy, Louisiana State
University, Baton Rouge, Louisiana 70803 USA}

\author{J.Y. Chan}
\affiliation{Department of Chemistry, Louisiana State
University, Baton Rouge, Louisiana 70803 USA}

\author{L.L. Henry}
\affiliation{Department of Physics, Southern University, Baton Rouge,
Louisiana, 70813 USA}

\author{J.F. DiTusa}
\affiliation{Department of Physics and Astronomy, Louisiana State
University, Baton Rouge, Louisiana 70803 USA}


\date{\today}

\begin{abstract}
The Hall effect and resistivity of the carrier doped magnetic
semiconductor Fe$_{1-x}$Co$_x$S$_2$ were measured for $0\le x \le
0.16$, temperatures between 0.05 and 300 K, and fields of up to 9
T. Our Hall data indicate electron charge carriers with a density of
only 10 to 30\% of the Co density of our crystals. Despite the
previous identification of magnetic Griffiths phase formation in the
magnetic and thermodynamic properties of this system for the same
range of $x$, we measure a temperature independent resistivity below
0.5 K indicating Fermi liquid-like transport. We also observe no
indication of quantum corrections to the conductivity despite the
small values of the product of the Fermi wave vector and the
mean-free-path, $1.5 \le k_F\ell \le 15$, over the range of $x$
investigated. This implies a large inelastic scattering rate such that
the necessary condition for the observation of quantum contributions
to the carrier transport, quantum coherence over times much longer
than the elastic scattering time, is not met in our samples. Above 0.5
K we observe a temperature and magnetic field dependent resistivity
that closely resembles a Kondo anomaly for $x$ less than that required
to form a long range magnetic state, $x_c$. For $x>x_c$, the
resistivity and magnetoresistance resemble that of a spin glass with a
reduction of the resistivity by as much as 35\% in 5 T fields. We also
observe an enhancement of the residual resistivity ratio by almost a
factor of 2 for samples with $x\sim x_c$ indicating temperature
dependent scattering mechanisms beyond simple carrier-phonon
scattering. We speculate that this enhancement is due to charge
carrier scattering from magnetic fluctuations which contribute to the
resistivity over a wide temperature range.

\end{abstract}


\maketitle

\section{Introduction and Motivation}

There has been enormous
interest\cite{custers,mathur,si,smith,doiron,stewart,schroder} over
the past decade in systems that can be tuned via pressure, magnetic
field, or composition, to be in proximity to zero temperature phase
transitions, or quantum critical points, QPC. This interest stems from
the non-standard, or non-Fermi liquid, NFL, behavior commonly
found\cite{custers,mathur,si,smith,doiron,stewart,schroder,takagi,kim}
in metals near QCP's.  Many of the systems investigated, particularly
those that are tuned by way of chemical substitution, are
significantly disordered and the role of the disorder in determining
the physical properties is not well
understood\cite{Griffiths,CastroNeto,castro2,vladrev,vojtarev,millis2}. Recent
theoretical work\cite{CastroNeto,castro2,vladrev,vojtarev,millis2} has
focused on the emergence of Griffiths phases in disordered materials
where statistically rare regions of order, or disorder, can dominate
the thermodynamic response. This appears particularly relevant near
zero temperature phase transitions where the formation of Griffiths
phases in disordered metals is thought to be the cause of the NFL
behavior observed over wide regions of composition, temperature and
magnetic field.

In a previous paper\cite{guo} we presented, and in the accompanying
article\cite{guoprb1} expand upon, the magnetic and thermodynamic
properties of the magnetic semiconductor Fe$_{1-x}$Co$_x$S$_2$ which
displays features consistent with the formation of Griffiths
phases\cite{vladrev,vojtarev}. We have demonstrated the formation of
clusters of magnetic moments both at low temperatures for $x$ less
than $x_c$, the critical concentration for a long ranged magnetic
ground state, as well as at temperatures above the critical
temperature, $T_c$, for $x>x_c$.  Here, $T_c$ is defined as the
temperature where the real part of the AC susceptibility displays a
maximum (see Ref.\ [\onlinecite{guoprb1}]).  In the temperature range
where these clusters form we find power-law temperature, $T$, and
magnetic field, $H$, dependencies with small exponents in agreement
with that predicted for Griffiths phases.

Here, we investigate in more full detail the charge transport
properties of these compounds to explore the consequences of Griffiths
phase formation. We find that the Co substitution induces metallic
behavior with a small density of electron-type carriers in all of our
Co doped crystals, including our most lightly doped ($x=3\times
10^{-4}$). For samples with $x\le x_c$ the resistivity, $\rho$, and
magnetoresistance, MR, are consistent with Kondo scattering of the
carriers\cite{kondo,hamann,schilling,schlottmann,andraka,rizzuto}. That
is, we observe a logarithmically decreasing $\rho(T)$ and a
$\rho(T,H)$ that scale as $(\rho - \rho(H=0))/\rho(H=0) = f[H/(T +
T^*)]$, where $T^*$ is the Kondo temperature and $\rho(H=0)$ the zero
field resistivity.  As we increase $x$ beyond $x_c$ we find that
magnetic ordering produces an increasing $\rho(T)$ up to temperatures
somewhat above $T_c$, similar to what is observed in metallic spin
glasses\cite{mydosh}. The application of a magnetic field of 5 T
dramatically reduces the resistivity of these samples by up to 35\% at
low temperatures. In this way the system resembles common metals with
magnetic impurities with Kondo behavior seen at low impurity
concentrations and a decreasing $\rho$ with decreasing $T$ below a
spin glass temperature for higher impurity
concentrations\cite{rizzuto,mydosh}.

As is true of nearly all metals that display Kondo anomalies, the
resistivity of our Fe$_{1-x}$Co$_x$S$_2$ crystals displays Fermi
liquid behavior at temperatures well below
$T^*$\cite{rizzuto,hamann,fisk}. Here, we observe a temperature
independent resistivity below 0.5 K for samples having both $x\le x_c$
as well as those with $x > x_c$. The main difference between our
crystals and common Kondo systems is that Fe$_{1-x}$Co$_x$S$_2$ has a
small density of charge carriers and poor conduction in the range of
$x$ that we investigate. A typical figure of merit for transport in
disordered metals is $k_F\ell$, where $k_F$ is the Fermi wave vector
and $\ell$ is the mean-free-path of the charge carriers. Here,
$k_F\ell$ is between 1.5 and 15 for $3\times 10^{-4} \le x\le 0.1$ a
range typical of doped semiconductors in proximity to the
insulator-to-metal (IM) transition where quantum contributions to the
conductivity are expected\cite{altshuler,lee}. Quantum interference
effects, such as the electron-electron interaction
effects\cite{altshuler,lee} described by Altshuler, are expected under
the conditions of diffusive transport and low temperatures such that
the elastic scattering time for carriers is much shorter than the time
required to randomize the charge carrier's quantum phase. Under these
conditions a square-root singularity in the electronic density of
states is thought to form, resulting in a $\sqrt{T}$ and $\sqrt{H}$
dependent
conductivity\cite{rosenbaum,rosenbaum2,sarachik}. Measurements in
doped semiconductors, including some magnetic semiconductors, display
the typical signatures of electron-electron interaction effects,
including the positive MR and a distinctive scaling of the
conductivity in $H/T$\cite{rosenbaum,sarachik,manyala}. We observe
none of these standard behaviors of disordered Fermi liquid behavior,
instead finding a negative MR and a temperature independent
conductivity at $T<$500 mK, despite the small values of $k_F\ell$ that
we infer.

The absence of quantum interference effects in the conductivity of our
samples is puzzling. We point out, however, that there is reason to
believe that the charge carriers have a substantial inelastic
scattering rate even at the lowest temperatures we measure. The high
density of low lying magnetic excitations associated with the magnetic
Griffiths phases that we have identified in this system are likely to
provide a large inelastic scattering cross-section for the charge
carriers, even at temperatures well below 1 K\cite{guo,guoprb1}. We
speculate that the condition for observation of the quantum
corrections to the conductivity, that the carriers retain quantum
coherence over times much longer than the inelastic scattering
time\cite{altshuler,lee}, is not met in our samples because of the
magnetic scattering. Reduction of the magnetic scattering by
application of magnetic fields as large as 9 T does not appear
sufficient to restore the quantum interference effects as the
resistivity continues to be $T$ independent even at such large fields.

Beyond the surprising lack of quantum interference effects, we observe
no indication of NFL behavior in the transport properties of our
Fe$_{1-x}$Co$_x$S$_2$ crystals. Thus, just as in the prototypical
semiconducting systems such as phosphorous doped
silicon\cite{paalanen,paalanen2,sarachik2}, NFL behavior observed in
the magnetic and thermodynamic properties that persist into the
metallic side of the IM transition, does not lead to NFL charge
transport. In Si:P, where disordered Fermi liquid transport is
observed\cite{paalanen}, the implication is that the conducting
electrons and the more localized dopants which dominate the
magnetization and specific heat, $C$, only interact weakly. It has
even been postulated that the these two electron fluids reside in
physically separate regions of the disordered semiconductor. In
contrast, in Fe$_{1-x}$Co$_x$S$_2$, we find convincing evidence that
the conducting electrons and the localized electrons responsible for
the magnetic properties of these materials interact substantially. The
Kondo resistance anomaly, so clearly defined in our data, indicates
the importance of such an interaction. Yet, surprisingly, we do not
observe any indication of NFL behavior in the charge carrier transport
to the lowest temperatures measured despite ample evidence for NFL
response of $M(H,T)$ and $C(H,T)$\cite{guo,guoprb1}.

There are several other peculiarities that we observe in our samples
with Co concentrations closest to the critical concentration for the
nucleation of a finite temperature long range magnetic phase. First,
we have measured a very large anomalous Hall coefficient, $R_S$, for
samples with $x$ near $x_c$. $R_S$ becomes much smaller as $x$ is
increased. The magnitude of $R_S$ does not appear to scale with the
resistivity of our samples in the manner predicted by the accepted
models\cite{luttinger,maranzana,jungwirth} of the anomalous Hall
effect. In addition, we observe a residual resistivity ratio, RRR,
defined as the resistivity at 300 K divided by the resistivity at 4 K,
that is nearly 2 for all samples except those with $x$ near $x_c$,
where it is almost twice as large. This indicates that there is an
additional temperature dependent scattering mechanism beyond the
typical phonon induced carrier scattering of common
metals\cite{zimon,gantmakher}. We demonstrate that fits of the Bloch
model for phonon scattering in metals do not completely describe
$\rho(T)$ for samples with $x\sim x_c$. Instead, we find that a simple
$AT^{\alpha}$ dependence with $\alpha \sim 1.5$, describes $\rho(T)$
well over a large temperature range. The success of this simple form
in modeling our data suggests that magnetic fluctuation scattering may
be important over a wide $T$ range for samples in proximity to a zero
temperature phase
transition\cite{custers,smith,doiron,schroder,takagi,nicklas,pfleiderer,moriya}.

The purpose of this article is to provide a more full rendering of our
Hall effect, resistivity, and magnetoresistance data and analysis than
that published previously\cite{guo}. In addition, we emphasize here
the absence of quantum contributions to the low temperature transport
that was not included in our previous work.

This paper is organized as follows: We outline some of important
experimental details of the sample preparation, the initial
characterization, and the measurement techniques in section
\ref{sec:exptdet}. This is followed by a presentation of our Hall
effect data and analysis. The resistivity and magnetoresistance of our
Fe$_{1-x}$Co$_x$S$_2$ crystals is presented in section
\ref{sec:conductivity}. Finally we summarize our results in section
\ref{sec:discuss}.

\section{Experimental Details \label{sec:exptdet}}

Single crystals of Fe$_{1-x}$Co$_x$S$_2$ were synthesized from high
purity starting materials, including Fe powder (Alfa Aesar 99.998\%),
Co powder (Alfa Aesar 99.998\%), and sulfur (Alfa Aesar 99.999\%) as
described previously\cite{guo,guoprb1}.  Initial characterization
included single crystal X-ray diffraction as described in
Refs.\onlinecite{guo,guoprb1}. The lattice constant determined by the
X-ray diffraction experiments\cite{guoprb1} showed a systematic
increase with Co concentration, $x$, in Fe$_{1-x}$Co$_x$S$_2$, beyond
the lattice constant for pure FeS$_2$, $a=0.54165$ nm and consistent
with the measurements of Ref.\onlinecite{bouchard}. The increase in
lattice constant with $x$ is consistent with Vegard's law and the idea
that Co replaces Fe within the pyrite crystal
structure. Energy-dispersive X-ray microanalysis (EDX) on JEOL
scanning electron microscope equipped with a Kevex Si(Li) detector was
performed to check the stoichiometry of our samples and were found to
be consistent with the magnetic moment density determined by our DC
magnetization measurements\cite{guoprb1}. Because magnetization
measurements were easily performed and highly reproducible, we have
subsequently used the saturated magnetization to determine the Co
concentration of our samples. Thus, throughout this manuscript the
stoichiometry of the samples noted in the figures and text was
determined in this manner. The variations in the saturated
magnetization for crystals from the same growth batch was measured to
be $\pm$ 10\% of the average value. Therefore, we report $x$
determined from measurements of $M$ at high field for the crystals
used in each of our measurements. Wherever possible, the same crystal
was employed for several different measurement types, particularly
for the Hall effect and magnetization measurements used to determine
anomalous Hall coefficients. As described in the accompanying article,
the AC susceptibility was used to establish the magnetic phase diagram
identifying Curie temperatures, $T_c$, and the critical Co
concentration for formation of a magnetic ground state, $x_c=0.007 \pm
0.002$\cite{guo,guoprb1}.

Resistivity, magnetoresistance (MR), and Hall effect measurements were
performed on single crystals polished with emery paper to an average
size of 0.5 mm x 1 mm x 0.1 mm. Thin Pt wires were attached to four
contacts made with Epotek conductive silver epoxy.  Hall effect
measurements were performed at 17 or 19 Hz on samples with carefully
aligned voltage leads in a Quantum Design MPMS gas flow cryostat from
1.8 to 300 K. These measurements were done in a superconducting magnet
with fields ranging from -5 to 5 T and the Hall voltage, $V_H$, was
determined as $V_H = [V(H) - V(-H)]/2$, to correct for any
contamination from the field symmetric MR due to misalignment of the
contacts.  The resistivity and MR measurements were performed at 17 or
19 Hz using standard lock-in techniques in the Quantum Design MPMS gas
flow cryostat with a 5 T superconducting magnet and a dilution
refrigerator equipped with a 9 T superconducting magnet.

\section{Experimental Results}

\subsection{Hall Effect
\label{halleffect}}
\begin{figure}[htb]
  \includegraphics[angle=90,width=3.0in,bb=85 395 545
  700,clip]{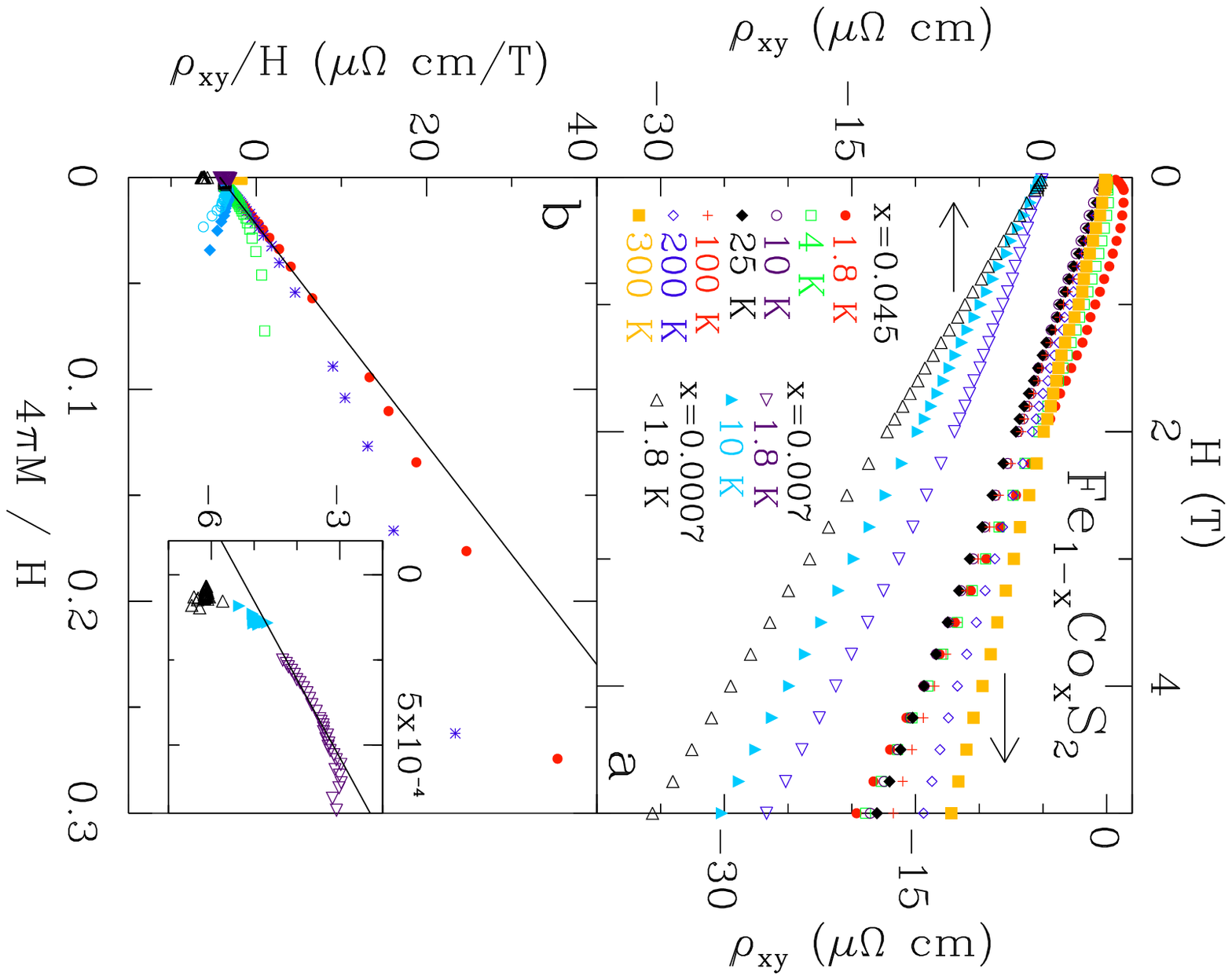}%
  \caption{\label{fig:hallpl} (Color online) Hall resistivity. a) The
Hall resistivity, $\rho_{xy}$, vs magnetic field, $H$, at several
temperatures, $T$, for 3 representative crystals with $T$'s and $x$'s
identified in the figure. Note that the $\rho_{xy}$ scale for the
$x=0.045$ sample is on the right-hand-side of the figure. b)
$\rho_{xy} / H$ plotted as a function of the magnetization, $M$,
divided by $H$ to highlight the anomalous part of the Hall
resistivity. Symbols the same as in frame a. The line is a fit of a
$\rho_{xy}/H = R_0 + R_S (4\pi M/H)$ form to the $x=0.045$ data at 1.8
K with $R_0$ and $R_S$ fitting parameters taken to be independent of
$H$. Inset: Same as frame b for $x=7\times 10^{-4}$ and $x=0.007$
samples on a smaller scale. Line is a linear fit, same form as in the
main frame of the figure, to the $x=0.007$ data at 4 K.}
\end{figure}

We begin our discussion of the transport properties of the charge
carriers donated by the Co substitution by presenting our Hall effect
measurements which give us an estimate of the charge carrier densities
in our crystals. In Fig.~\ref{fig:hallpl}a the Hall resistivity,
$\rho_{xy}$, is plotted for 3 representative samples including our
$x=0.0007$ sample (at 1.8 K), our $x=0.007$ sample (at 1.8 and 10 K),
and our $x=0.045$ sample at a series of temperatures between 1.8 and
300 K. All samples display a negative $\rho_{xy}$, indicating n-type
carriers, except at low fields and temperatures where a positive Hall
potential is found for samples with $x \ge x_c$. This large positive
contribution to $\rho_{xy}$ at low fields, demonstrated in the figure
for our $x=0.007$ and $x=0.045$ samples, is suppressed with
temperature so that by warming to 10 K we again find a negative
$\rho_{xy}(H)$ which is linear in $H$. The observation of a low-field
positive contribution to $\rho_{xy}$ is not surprising since magnetic
materials typically have two contributions to their Hall
resistivities, an ordinary part due to the Lorentz force experienced
by the carriers proportional to $H$ and inversely
proportional to the carrier density, and an anomalous part
proportional to the sample's magnetization. It is common to
parameterize $\rho_{xy}$ as
\begin{equation} 
\rho_{xy} = R_0H + 4\pi M R_S
\label{eq:anhall}\end{equation} 
with $R_0$ the ordinary Hall coefficient, $R_S$ the anomalous Hall
coefficient, and $M$ the magnetization, to highlight these
contributions\cite{luttinger,maranzana,jungwirth}. Interest in the
anomalous Hall effect has grown over the past few years because of the
large contribution it makes to $\rho_{xy}$ in magnetic
semiconductors\cite{jungwirth,manyala2,ohno1,ohno2,ohno3}. The present
understanding of the anomalous Hall effect includes contributions from
extrinsic sources, due to spin-orbit scattering, and intrinsic
sources, from spin-orbit effects inherent in the material's band
structure\cite{luttinger,jungwirth}. These theories predict a strong
dependence of $R_S$ on the carrier scattering rate such that $R_S
\propto \rho_{xx}^2$ for intrinsic or side jump scattering and
$R_S\propto \rho_{xx}$ for skew scattering dominated
transport\cite{maranzana,jungwirth}.

As is standard practice, we display in Fig.~\ref{fig:hallpl}b the
quantity $\rho_{xy}/H$ as a function of $4\pi M / H$ where $M$ is the
measured magnetization of these same crystals. Plotting the data in
this way allows a comparison of our measured $\rho_{xy}$ with
Eq.~\ref{eq:anhall}. In the case where the ordinary Hall coefficient
is field independent, as it is for a single carrier band, a linear
$M/H$ dependence is often found. Here we see significant curvature to
$\rho_{xy}$ indicating that there are either multi-band effects in
$R_0$, or that there is a field dependence to the anomalous
coefficient. Field dependent $R_S$ values are observed in materials
having large MRs since the carrier scattering rates are field
dependent\cite{mlee}. However, such an analysis was not successful in
modeling the $M/H$ dependence of $\rho_{xy}/H$ shown in
Fig.~\ref{fig:hallpl}b. A significant temperature dependence of $R_S$
is also apparent for our $x=0.045$ sample in Fig.~\ref{fig:hallpl}b.

\begin{figure}[htb]
  \includegraphics[angle=90,width=3.0in,bb=60 380 550
  700,clip]{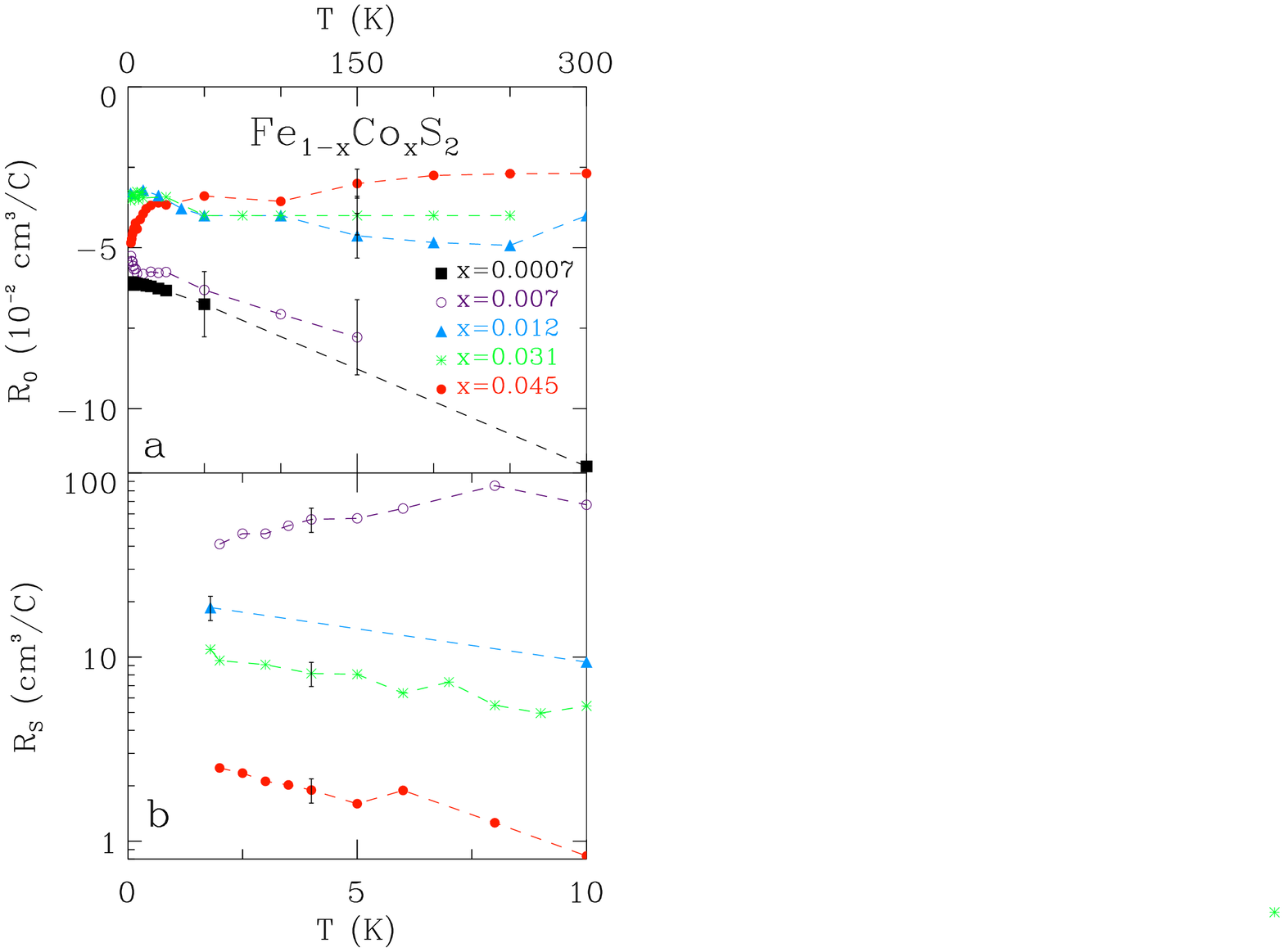}%
  \caption{\label{fig:hall2pl} (Color online) Temperature dependence
of the Hall coefficients. a) The ordinary Hall coefficient, $R_0$, as
a function of temperature, $T$, for five representative crystals with
$x$ identified in the figure. $R_0$ is determined from fits of the
Hall resistivity to the standard form $\rho_{xy} = R_0 H + 4\pi M R_S$
where $M$ is the measured magnetization and $R_S$ is the anomalous
Hall coefficient. b) Temperature dependence of $R_S$ for 4 crystals
identified in frame a. $R_S$ determined with the same procedure as in
frame a.  }
\end{figure}

The results of parameterizing $\rho_{xy}$ as in Eq.~\ref{eq:anhall}
are shown in Fig.~\ref{fig:hall2pl} where $R_0$ and $R_S$ are plotted
as a function of $T$ for five samples including the same three samples
displayed in Fig.~\ref{fig:hallpl}. $R_S$ for our $x=0.0007$ crystal
was omitted from the figure because $\rho_{xy}$ is highly linear and
the magnetization of this sample too small to allow an accurate
determination. The $T$-dependence of both $R_0$ and $R_S$ that were
apparent in Fig.~\ref{fig:hallpl}a and b are made quantitative with
the use of this simple form for the Hall resistivity. What is
interesting, first, is the observation of significant temperature
dependencies of the Hall coefficients below 10 K where the magnetic
susceptibility, specific heat, and, as we demonstrate below, $\rho$
all display unusual behavior associated with the magnetic properties
of these materials\cite{guo,guoprb1}. Second, we find very large $R_S$
values for our samples that are very close to the critical point for
magnetic ordering.

\begin{figure}[htb]
  \includegraphics[angle=90,width=3.0in,bb=70 385 520
  705,clip]{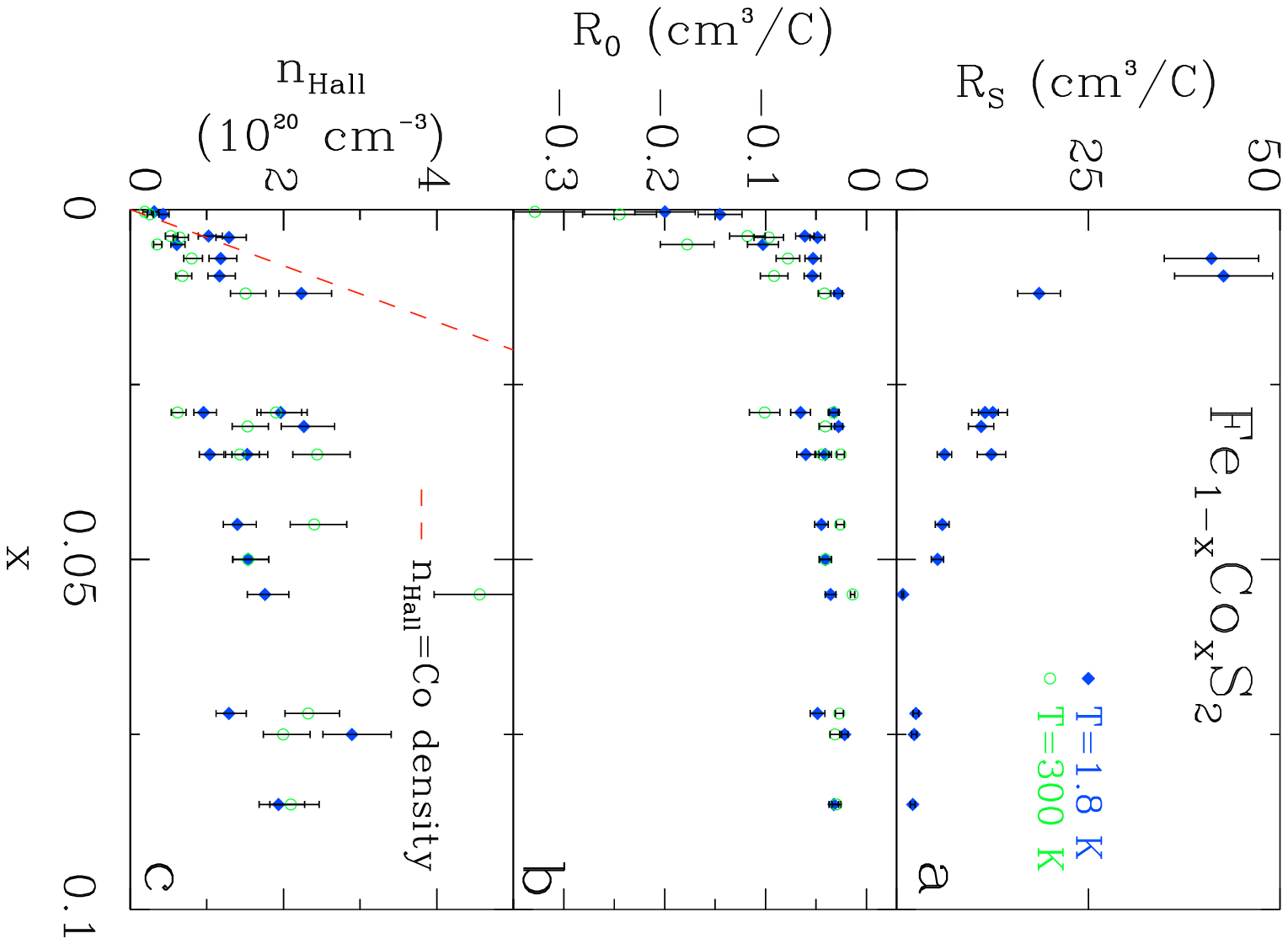}%
  \caption{\label{fig:hall3pl} (Color online) Cobalt concentration
dependence of the Hall coefficients and carrier density.  a) Cobalt
concentration, $x$, dependence of the anomalous Hall coefficient,
$R_S$ at 1.8 K. $R_S$ was determined from fits of the form $\rho_{xy}
= R_0 H + 4\pi M R_S$ to the Hall resistivity, $\rho_{xy}$, where
$R_0$ is the ordinary Hall coefficient. b) $x$ dependence of $R_0$
determined from the same fits at 1.8 and 300 K as identified in frame
a.  c) Carrier concentration, $n_{Hall}$ determined from the simple
form for the ordinary Hall coefficient, $R_0 = 1/n_{Hall} ec$, where
$e$ is the electronic charge and $c$ is the speed of light, at 1.8 and
300K as identified in frame a. Dashed red line represents the carrier
concentration expected if each Co dopant were to donate a single
electron to a conducting band.  }
\end{figure}

The $x$ dependence of $R_S$ and $R_0$ for a large number of crystals
is presented in Fig. \ref{fig:hall3pl}.  Here we demonstrate the very
large values of $R_S$ found near $x_c$ along with a continuous
decrease with $x$ beyond $x_c$. We note that neither the standard
extrinsic nor intrinsic theories of the anomalous Hall coefficient can
explain the $x$ dependence of $R_S$ that we measure here by way of a
scattering rate variation with $x$. As we demonstrate in the
discussion of the conductivity below, we measure very small variation
in $\rho$ over the range in $x$ where the large decrease in $R_S$ is
apparent in Fig.~\ref{fig:hall3pl}a. At present, we do not have an
adequate understanding of the very large values of $R_S$ we measure
near $x_c$, although we speculate that an inhomogeneous magnetic
state, the Griffiths phase, for $x\sim x_c$ identified from our
magnetic and thermodynamic measurements, may amplify the anomalous
Hall effect near $x_c$\cite{guo,guoprb1}. In frame c of the figure we
plot the carrier density calculated from the simple form $R_0 =
1/n_{Hall}ec$, where $n_{Hall}$ is the carrier density, $e$ is the
electronic charge, and $c$ is the speed of light. We observe that
$n_{Hall}$ increases with Co substitution although it is much smaller
than the Co density of these crystals, particularly for $x>0.025$. Our
values of $n_{Hall}$ indicate a carrier concentration of only 10 to
30\% of the Co density. Thus, there appears to be significant fraction
of the electrons added by the Co substitution that are in localized
states.

Our Hall data have established the sign of the charge carriers,
negative as expected for Co substitution in FeS$_2$, and that the
carrier density is only 10 to 30\% of the Co density of our crystals.
In addition, we find very large Hall conductivities resulting from
extraordinarily large anomalous Hall coefficients for samples close to
the critical concentration for magnetism.

\subsection{Conductivity \label{sec:conductivity}}
\begin{figure}[htb]
  \includegraphics[angle=90,width=3.0in,bb=60 300 540 710,
clip]{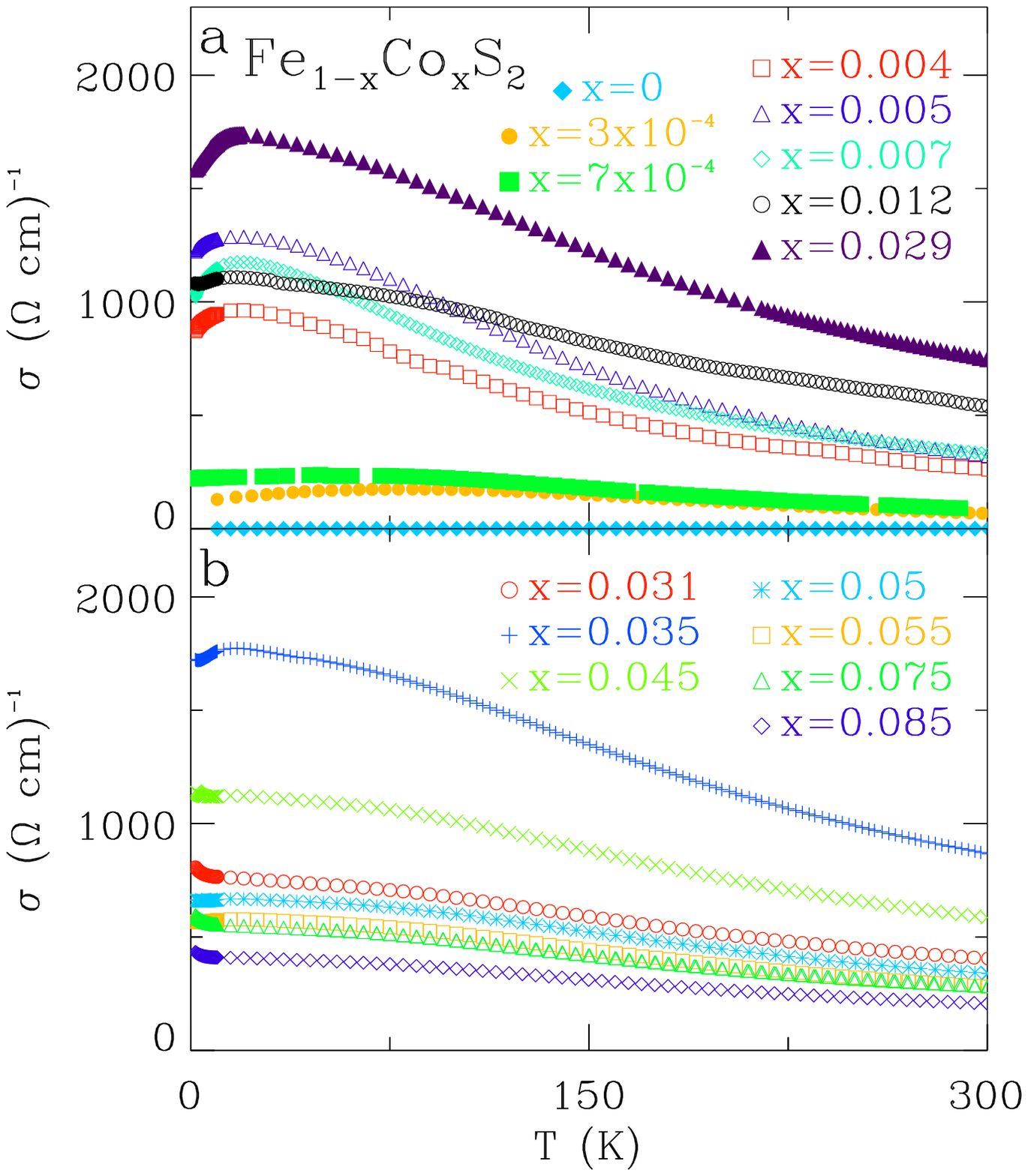}%
  \caption{\label{fig:sigpl} (Color online) Conductivity.  a) and b)
Temperature, $T$, and cobalt concentration, $x$, dependence of the
electrical conductivity, $\sigma$, of several representative
crystals. The stoichiometry of the crystals is identified in the
figure.}
\end{figure}

\begin{figure}[htb]
  \includegraphics[angle=90,width=3.0in,bb=60 110 540 710,
clip]{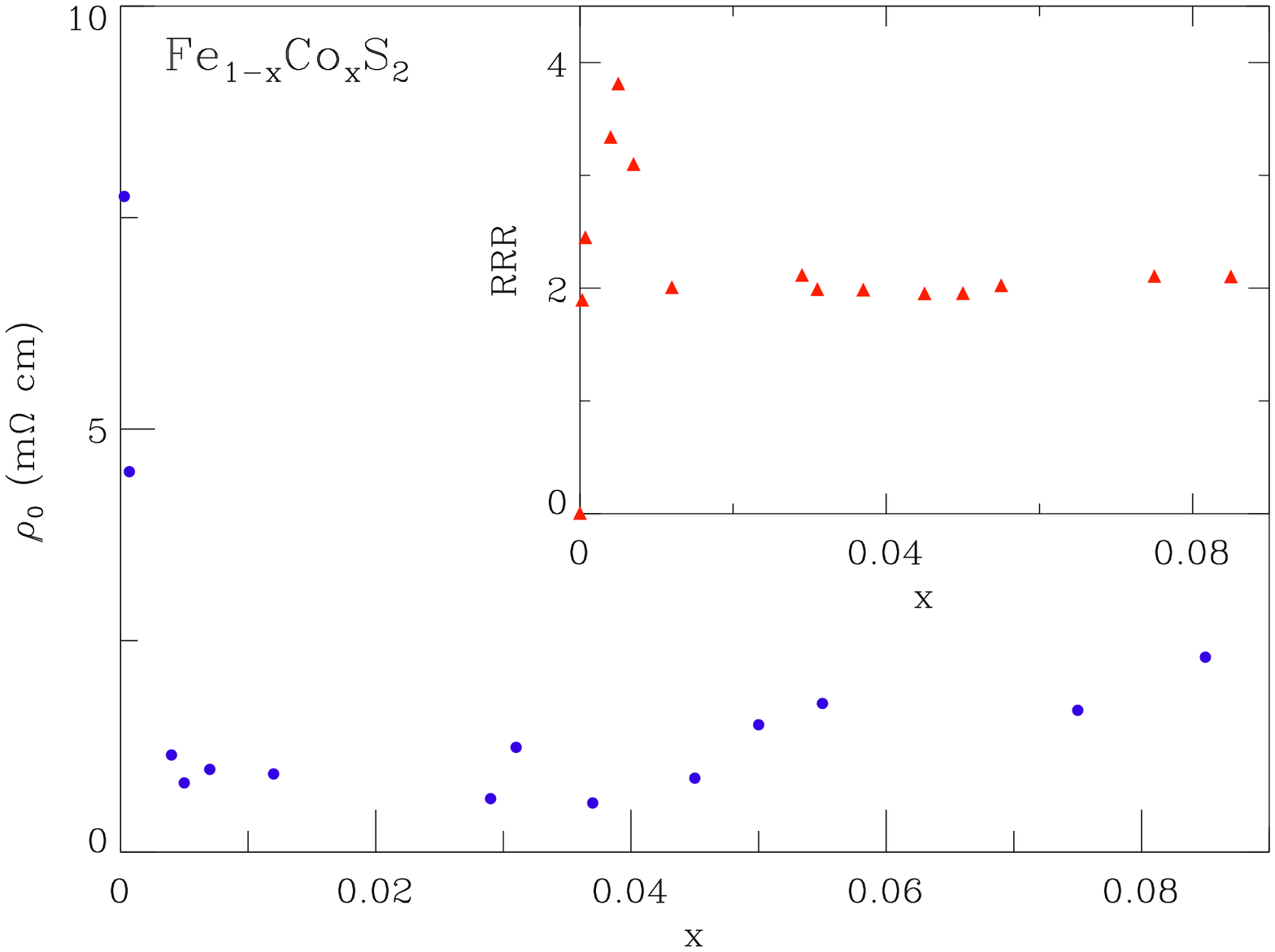}%
  \caption{\label{fig:resrho} (Color online) Residual resistivity and
residual resistivity ratio. The residual resistivity, $\rho_0$,
(resistivity at 1.8 K) of our Fe$_{1-x}$Co$_x$S$_2$ crystals. Inset;
Residual resistivity ratio, RRR, defined as the resistivity at 300 K
divided by $\rho_0$}
\end{figure}

After establishing that our crystals were single phase and that Co
successfully replaces Fe in FeS$_2$ adding electron-like carriers, we
determined the conductivity $\sigma$ of our samples.
Fig. \ref{fig:sigpl} shows that, although the nominally pure FeS$_2$
crystal is insulating, all of our crystals with Co substitutions were
metallic having a conductivity that extrapolates to a non-zero value
at $T=0$\cite{jarrett}. Further, our crystals have $\sigma$ that
increases with $x$ for $x \le 0.035$ and decrease for larger
$x$. These trends are made clearer in Fig.~\ref{fig:resrho} where we
plot the residual resistivity defined as $\rho_0 = \sigma^{-1}$ at
$T=1.8$ K. Since the Hall effect measurements presented in section
\ref{halleffect} display a continuous increase in $n_{Hall}$ with $x$
for $x<0.025$ followed by a relatively constant carrier density at
larger $x$, the decreased $\sigma$ at $x>0.035$ indicates that the Co
impurities, or other substitution dependent defects, represent strong
scattering centers for the carriers. The conductivity decreases with
$T$ over most of the temperature range also indicating metallicity for
all $x>0$. We summarize the $T$-dependence of $\sigma$ by plotting the
residual resistivity ratio, RRR, of our crystals in the inset to
Fig.~\ref{fig:resrho}. Here RRR is defined as ratio of the
resistivity, $\rho = 1/\sigma$, at 300 K to that at 4 K and the figure
demonstrates that RRR is shown to be between 1.9 and 4 for all of our
$x>0$ samples. Metals with small disorder typically display a large
RRR due to a strongly $T$-dependent scattering rate for charge
carrier-phonon scattering below the Debye temperature ($\Theta_D =
610$ K for FeS$_2$)\cite{robie,kansy}. For disordered metals this
dependence can be hidden by a large impurity, $T$-independent,
scattering rate. In our samples, the RRR is small, a second indication
that the impurities and defects related to the chemical substitution
represent strong scattering centers for the doped carriers. In
addition to the temperature independent impurity related scattering
and the $T$-dependent phonon scattering apparent in the conductivity,
there appears to be a separate, $T$-dependent, contribution to the
conductivity of our crystals evident below $\sim 20$ K.

\subsubsection{Resistivity Below 20 K}
\begin{figure}[htb]
  \includegraphics[angle=90,width=3.5in,bb=60 50 520 735,
clip]{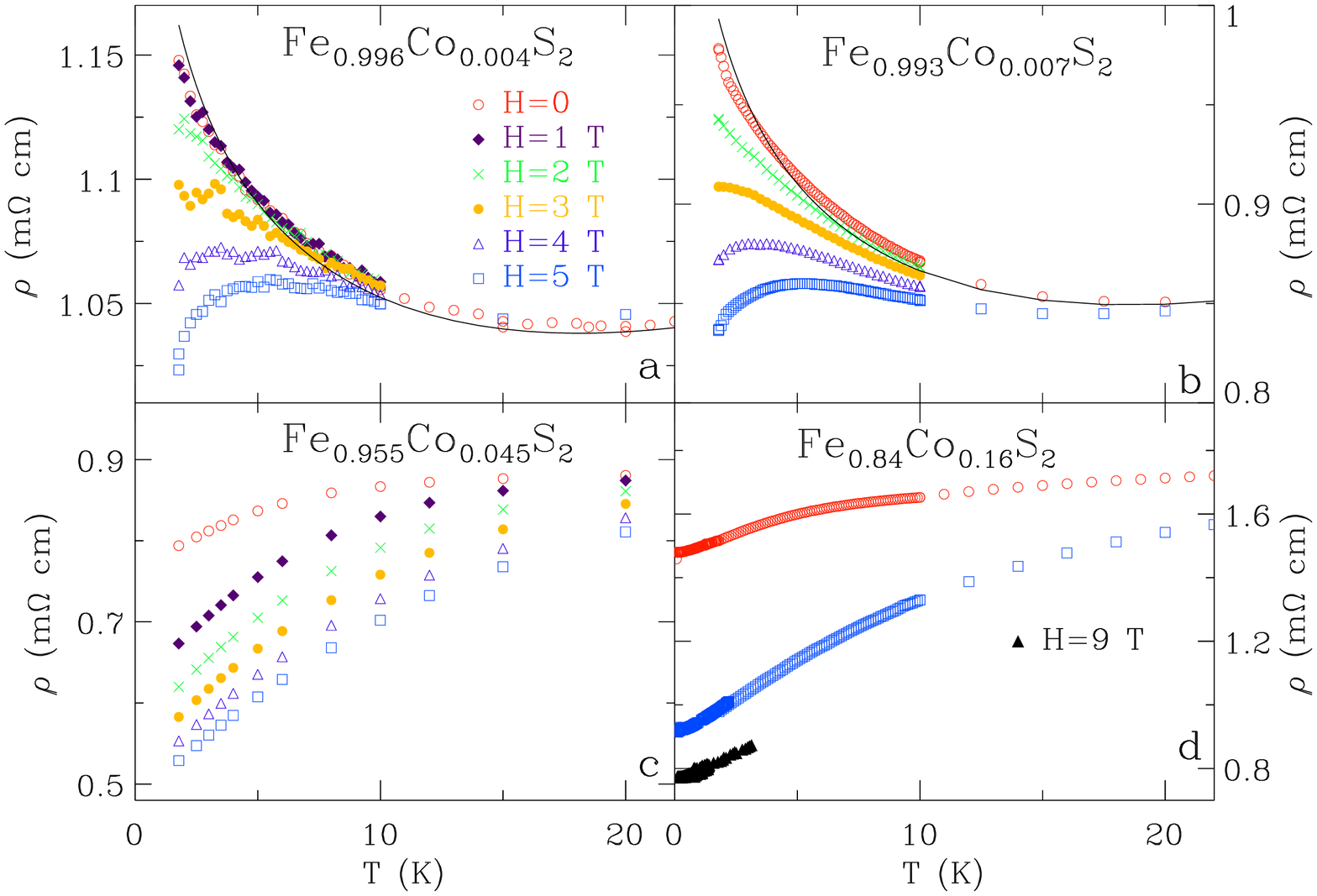}%
  \caption{\label{fig:mrpl} (Color online) Resistivity and
Magnetoresistance.  a) through d) Temperature, $T$, dependence of the
resistivity, $\rho$ of four single crystals at magnetic fields, $H$,
and stoichiometry's identified in the figure. Zero field data same as
in Fig.\ {\protect{\ref{fig:sigpl}}}. Solid line is a fit of a Kondo
anomaly form to the data using the Kondo temperature determined from a
scaling of the magnetoresistance data (see
text){\protect{\cite{hamann}}}.}
\end{figure}

\begin{figure}[htb]
  \includegraphics[angle=90,width=3.0in,bb=75 120 510
  710,clip]{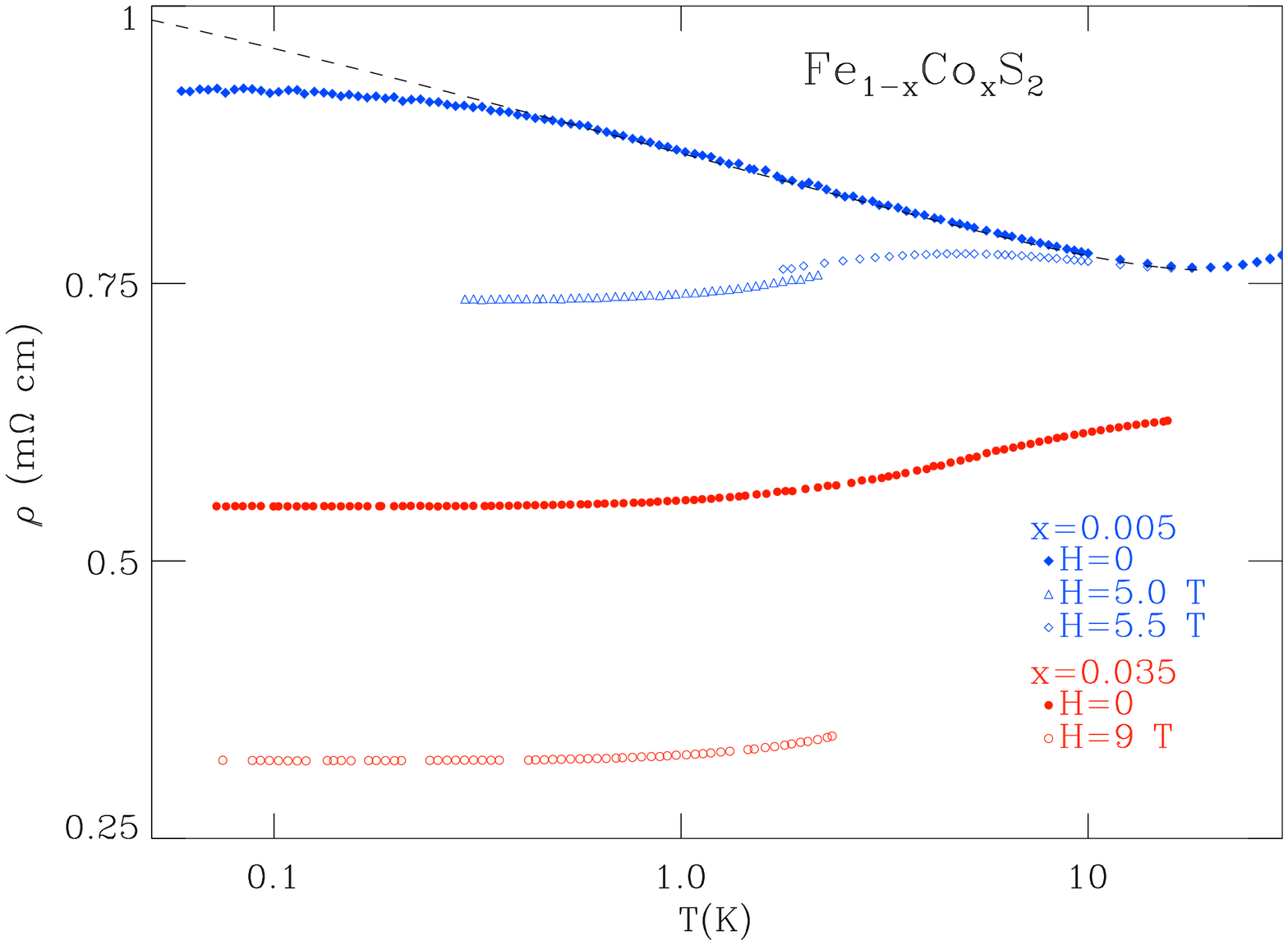}%
  \caption{\label{fig:rholtpl} (Color online) Low temperature
resistivity.  The resistivity, $\rho$, as a function of temperature,
$T$, at $0.07 <T<20$ K for two representative
crystals. Stoichiometry and magnetic fields are identified in the
figure. Dashed line is a fit of the Kondo
form{\protect{\cite{hamann}}} to the data for a $x=0.005$ crystal
above 0.5 K with the Kondo temperature held at 1.4 K as determined
from the best scaling of the magnetoresistance data. See text for
details.  }
\end{figure}

The temperature dependence of the charge carrier transport at low
temperatures can often reveal much about the character of the charge
carriers and their scattering. Therefore we have explored the carrier
transport below 20 K in some detail as shown in Figs. \ref{fig:mrpl}
and \ref{fig:rholtpl} where we display the resistivity of a few
representative crystals.  As these figures demonstrate, crystals with
$x < 0.01$ generally display a decreasing $\rho$ with $T$ below 20 K,
while those with $x>0.01$ display an increasing $\rho$ with $T$
leading to a maximum in $\rho$ at temperatures somewhat above
$T_c$\cite{guoprb1}.  In addition, we observe that $\rho$ tends toward
saturation by $T<0.3$ K, a behavior typically observed in metals
without substantial disorder where it is indicative of transport in a
Fermi liquid.  The resistivities we measure are above 500 $\mu\Omega$
cm, consistent with the small carrier concentration revealed in our
Hall effect measurements and a large scattering rate. A typical
measure of the proximity to the insulator-to-metal transition is the
quantity $k_F\ell$, where $k_F$ is the Fermi wave vector and $\ell$ is
the mean-free-path of the carriers. Our estimates of $k_F\ell$ at 4 K
for the samples shown in Fig.~\ref{fig:sigpl} range from 1.5 for our
smallest $x$, to 15 for the crystals having the smallest $\rho_0$ in
Fig.~\ref{fig:resrho}. With $k_F\ell$ being close to 1, it is
generally expected that $\rho$ display evidence of electron-electron
interaction effects that dominate the low-$T$ magnetotransport of
prototypical semiconductors\cite{altshuler,lee,rosenbaum}, and some
magnetic semiconductors\cite{manyala,manyala3}, near the IM
transition. The conductivity of common semiconductors with doping
concentrations near that required for a IM transition displays both a
$T^{1/2}$ dependence at low-$T$ and a positive MR that grows as
$H^{1/2}$ for $g\mu_BH > k_B T$. We show below that we observe neither
of these signatures of disordered Fermi-liquid transport in
Fe$_{1-x}$Co$_x$S$_2$. In this same range of $x$ and $T$, our specific
heat and magnetic susceptibility measurements reveal the formation of
Griffiths phases which have been suggested to be a cause of non-Fermi
liquid behavior\cite{guo,guoprb1,CastroNeto,vladrev}. Instead, we
observe a temperature independent resistivity below 1 K with no
indication of quantum interference effects in the carrier
transport. The coincidental observation of the formation of an
interesting disordered magnetic phase and the absence of quantum
interference in the electrical conductivity suggests that the
scattering of carriers from magnetic excitations associated with the
Co dopant ions is substantially inelastic. This large inelastic
scattering rate may be sufficient to cut off any quantum interference
effects associated with the diffusive transport of the carriers.

As is apparent in Figs.~\ref{fig:mrpl} and \ref{fig:rholtpl} the
application of magnetic fields substantially decreases $\rho$ for all
samples measured, independent of the sign of $d\rho/dT$. The decrease
in resistivity is as large as 35\% of $\rho_0$ for a 5 T field. The
negative MR we observe is at odds with what has been measured in
common semiconductors such as Si:P, and other magnetic semiconductors,
such as Fe$_{1-x}$Co$_x$Si, where a positive MR is associated with the
electron-electron interaction
effects\cite{rosenbaum,sarachik,manyala}. For samples that display a
$d\rho/dT < 0$ at $H=0$ the effect of magnetic field is is to reverse
the sign of the temperature derivative so that all samples have a
$d\rho / dT>0$ at high fields. Instead of resembling the transport of
charge carriers in prototypical semiconductors, we observe a behavior
similar to what is seen when magnetic impurities are added to high
purity metals\cite{rizzuto}.

In common metals the addition of magnetic impurities leads to a low
temperature resistivity anomaly that has been extensively
investigated\cite{kondo,rizzuto} for over 40 years. Small impurity
densities, typically less than 0.01\%, induce an increased resistivity
with decreasing $T$ similar to what we observe here. This is the well
known Kondo effect\cite{kondo,rizzuto} where conduction electrons
increasingly screen the magnetic impurities as the temperature is
lowered. The effect of a magnetic field is to Zeeman split the energy
levels of the impurity moments effectively removing the Kondo
resonance, thereby removing a scattering channel for the carriers. At
higher impurity concentrations, interactions between impurity magnetic
moments become of the same order as the effective screening
temperature, $k_BT^*$, where $T^*$ is the Kondo temperature and $k_B$
is Boltzmann's constant. In this case, a spin glass or a disordered
magnetic state forms at low-$T$ and is typically indicated by a
decrease in $\rho$ below the glass freezing or magnetic ordering
temperature\cite{mydosh}. We note that the $T$ and $x$ ranges where we
observe $d\rho/dT > 0$ at $H=0$ correspond well with the magnetically
ordered states identified by a peak in the AC
susceptibility\cite{guoprb1}.

In Figs. \ref{fig:mrpl} and \ref{fig:rholtpl} we have included a
comparison of our data for $\rho(T)$ at $H=0$ with a standard form for
the Kondo resistance anomaly given by Hamann\cite{hamann,schilling}
\begin{equation} 
\rho_{spin} = S\rho_0 \left(1 - \cos{2\delta_{\nu}}{{\ln{T/T_K}}
\over {(\ln^2{T/T_K} + \pi^2S(S+1))^{1/2}}}\right),
\label{eq:hamann}\end{equation} 
where $T_K$ is the Kondo temperature, $S$ the spin of the impurity,
$\delta_{\nu}$ the phase shift due to ordinary scattering, and
$\rho_0$ the the s-wave unitarity limit resistivity. We have fit this
form to our data for $x<x_c$ with $S$ set at 1/2 and $T_K$ held at the
value of $T^*$ found from the scaling of the MR, see below, and found
best fit values for $\delta_{\nu}$ between 25 and 60$^o$.

\begin{figure}[htb]
  \includegraphics[angle=90,width=3.0in,bb=60 60 510
  740,clip]{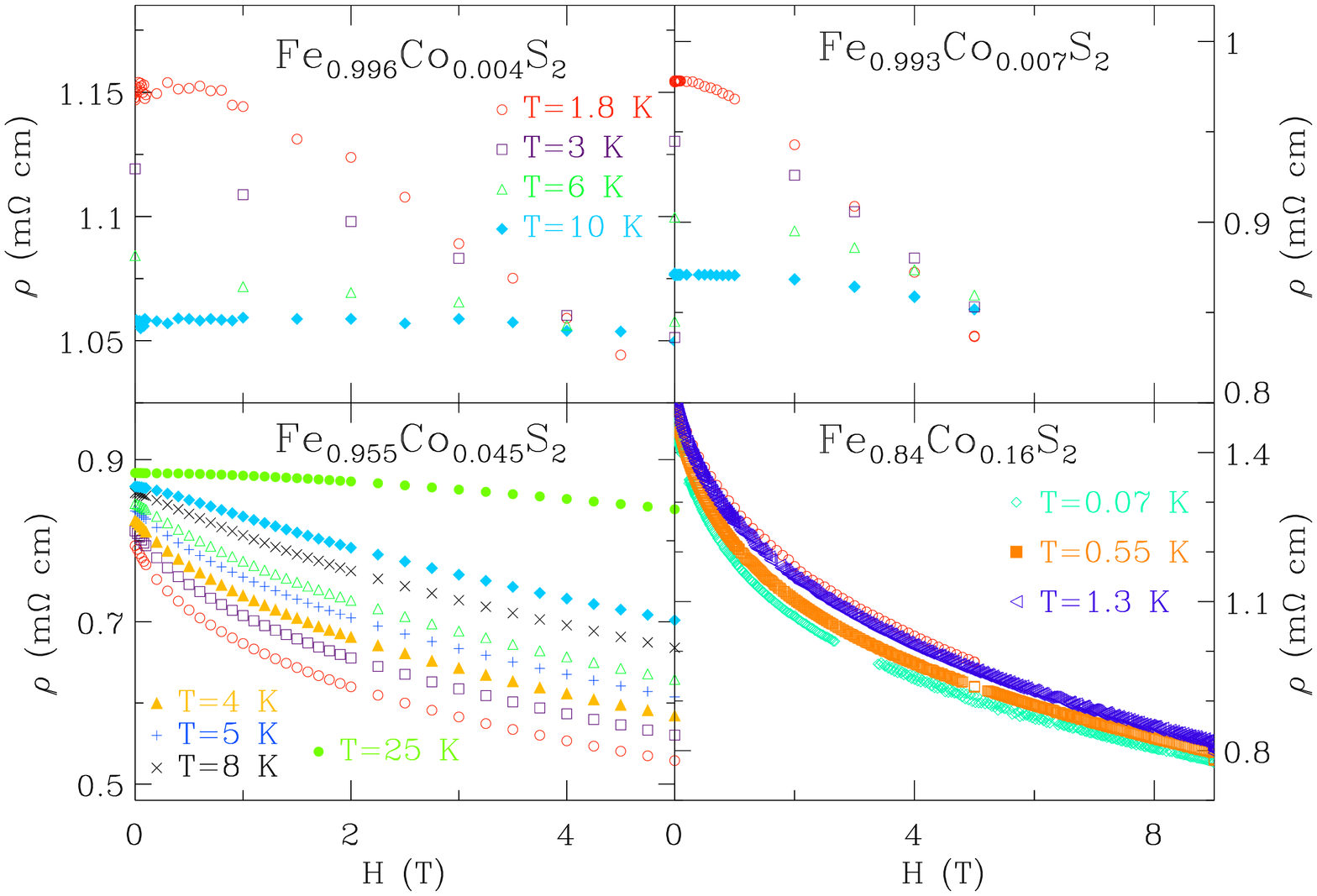}%
  \caption{\label{fig:mrfspl} (Color online) Magnetoresistance. a)
through d) The magnetoresistance (resistivity, $\rho$ as a
function of magnetic field, $H$) for the same crystals as in Fig.\
{\protect{\ref{fig:mrpl}}} with stoichiometry's and temperatures, $T$,
identified in the figure. The current direction was perpendicular 
to  $H$, transverse MR, for all data presented in the figure. }
\end{figure}

The field dependence of $\rho$ at constant $T$ for the same 4 crystals
as in Fig.~\ref{fig:mrpl} is shown in Fig. \ref{fig:mrfspl} where the
negative MR is once again displayed. The size of the MR decreases with
temperature so that by about $T=25$ K there is only a small MR ($<
5$\%).  The MR at 1.8 K is large indicating that the scattering of the
carriers by the magnetic impurities is dominating the resistivity of
our crystals, particularly at larger $x$. We also observe a change in
the shape of the MR at low $H$ for $x > 0.01$. Although the MR is
negative and analytic at $H=0$ for samples with $x<x_c = 0.007\pm
0.002$, as well as for $x>x_c$ at $T>T_c$, it is very sharp at
$T<T_c$, a feature commonly observed in ferromagnetic
metals\cite{campbell}.

\begin{figure}[htb]
  \includegraphics[angle=90,width=3.0in,bb=260 320 540
  710,clip]{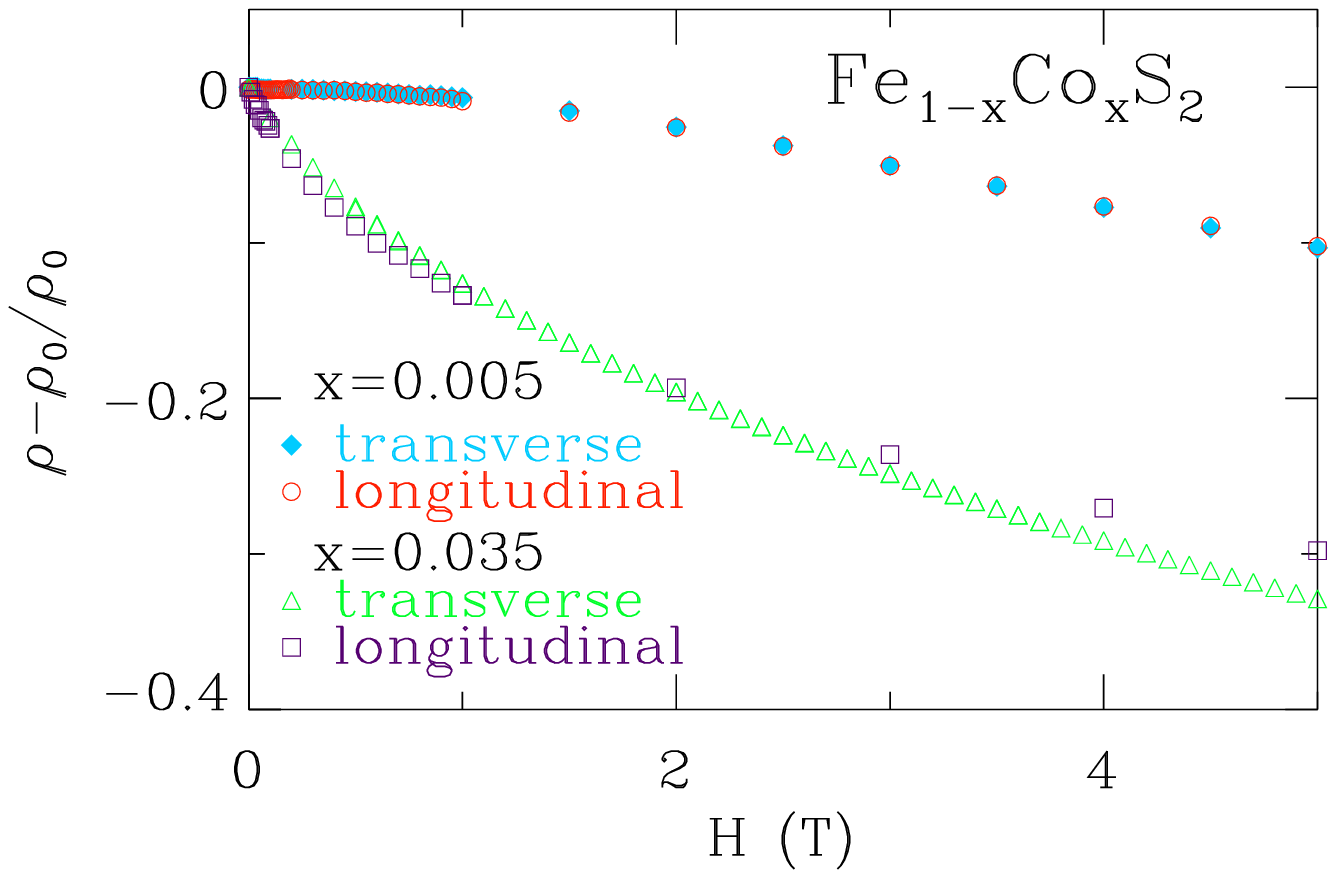}%
  \caption{\label{fig:mrtlpl} (Color online) Transverse and
longitudinal magnetoresistance. a) The magnetoresistance, MR, defined
as the resistivity, $\rho$, as a function of magnetic field, $H$ after
subtracting the resistivity at $H=0$, $\rho_0$, and dividing by
$\rho_0$, $\rho - \rho_0 / \rho_0$, of two representative crystals at
1.8 K. Stoichiometry's of the crystals are identified in the
figure. The longitudinal MR refers to the current direction being
parallel to $H$. The transverse MR refers to the measurement geometry
where the current is in the same direction with respect to the crystal
as in the longitudinal MR measurement, but with the crystal oriented
such that the current is perpendicular to $H$.  }
\end{figure}

A common indicator for the mechanism of the MR is the difference in MR
for currents parallel and perpendicular to the magnetic field. In
Fig. \ref{fig:mrtlpl} we plot the MR measured for two representative
samples in the two current-field configurations. We have chosen one
sample with $x<x_c$ and one with $x>x_c$ for this plot, and in both
cases there is little change in the MR with field orientation. We
conclude that the contributions to the MR from orbital effects is
minimal in agreement with our tentative assignment of the MR, and
$T$-dependent $\rho$, to a Kondo anomaly.

\begin{figure}[htb]
  \includegraphics[angle=90,width=3.2in,bb=60 60 520
  730,clip]{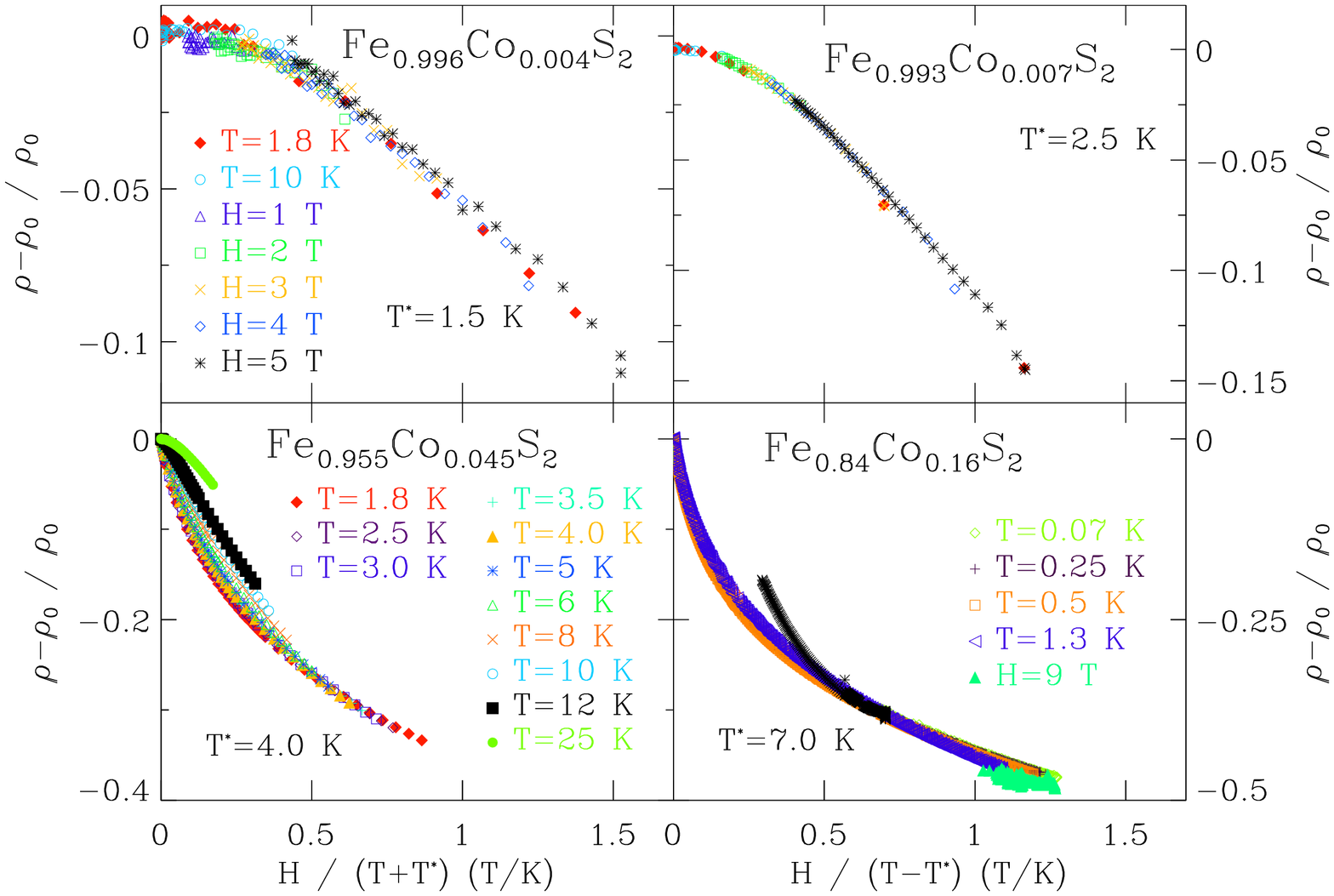}%
  \caption{\label{fig:mrscapl} (Color online) Scaling plot of the
magnetoresistance. a) through d) Magnetoresistance, $\rho - \rho_0 /
\rho_0$ where $\rho$ is the resistivity and $\rho_0$ refers to the
resistivity in zero magnetic field, $H$, as a function of $H$ divided
by the temperature, $T$, added to the Kondo temperature, $T^*$
determined by the best scaling of the these data. The stoichiometry of
the crystals is indicated in the figure, as is the value of $T^*$
which leads to the best scaling of the data. $T$ and $H$ for the
constant temperature and field scans are indicated in the figure.  The
scaling is seen to work well for samples where $x < x_c$, the critical
Co concentration for ferromagnetism, frames a and b. In frames c and d
where $x>x_c$, no value of $T^*$ results in reasonable scaling of all
the data.  }
\end{figure}

\begin{figure}[htb]
  \includegraphics[angle=90,width=3.0in,bb=85 110 520
  680,clip]{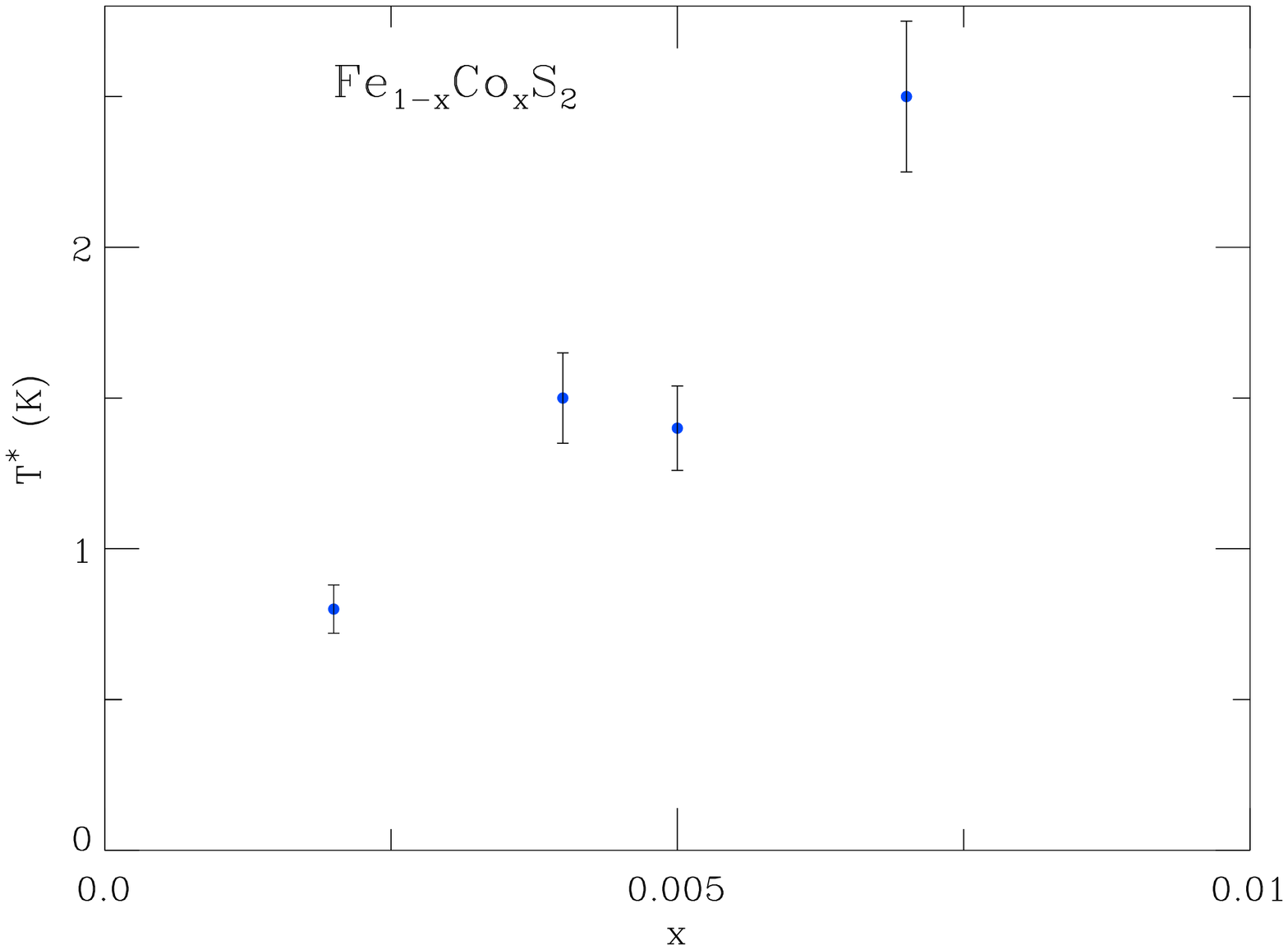}%
  \caption{\label{fig:tkpl} (Color online) Kondo temperature. The
Co concentration dependence, $x$ of the Kondo temperature $T^*$ as
determined from scaling of the magnetoresistance, see Fig.\
{\protect{\ref{fig:mrscapl}}} for example. Only samples where $x$ is
less than the critical concentration for ferromagnetism, $x_c$, is
$T^*$ accurately determined and only those data are plotted in the
figure.  }
\end{figure}

To further test the hypothesis that the MR is related to scattering
from magnetic moments associated with the Co substitution, and more
specifically the Kondo effect for $x<x_c$, we have attempted a simple
scaling of the MR to a standard Kondo form, $(\rho - \rho_0) / \rho_0
= f(H/(T+T^*)$\cite{schlottmann,andraka}. In Fig. \ref{fig:mrscapl} we
display the typical results of our scaling procedure for crystals with
$0.004\le x\le 0.16$. While the scaling quality is within the scatter
of the data for paramagnetic samples, $x<x_c=0.007$, there is
significant deviation from scaling for samples having a magnetic
transition at finite $T$. The values for $T^*$ resulting in the best
scaling of our data are plotted in Fig. \ref{fig:tkpl} where $T^*$ is
seen to increase with $x$.  We conclude from the quality of the
scaling of the data that a single ion Kondo form does well to describe
the resistivity where the interaction between moments is expected to
be small, that is, for $x\le x_c$ and $T$ larger than a scale related
to the interactions between local moments, such as an RKKY energy
scale. This is in accord with our magnetization measurements for
samples with $x<x_c$ where the $T$ dependence of the magnetic
susceptibility displays a small negative Weiss
$T$\cite{guo,guoprb1}. From the quality of the fits shown in the
figures, and the quality of the scaling of the MR in
Fig.~\ref{fig:mrscapl}, we conclude that a single ion Kondo form
describes the low temperature transport of our samples with $x<x_c$
well, while for $x>x_c$ the transport properties resemble those of
Kondo systems where the concentration of magnetic impurities is large
enough so that interaction between them results in magnetic ordering.

\subsubsection{Resistivity above 20 K}
Above 20 K the conductivity shown in Fig. \ref{fig:sigpl} decreases
with increased $T$ demonstrating metallic conduction of the carriers
introduced by Co substitution. However, the thermally induced
scattering does not decrease the conductivity of our crystals by more
than a factor of 2 in most crystals (see RRR in Fig.~\ref{fig:resrho})
reflecting the importance of the substitutional disorder.

\begin{figure}[htb]
  \includegraphics[angle=90,width=3.2in,bb=60 85 550
  685,clip]{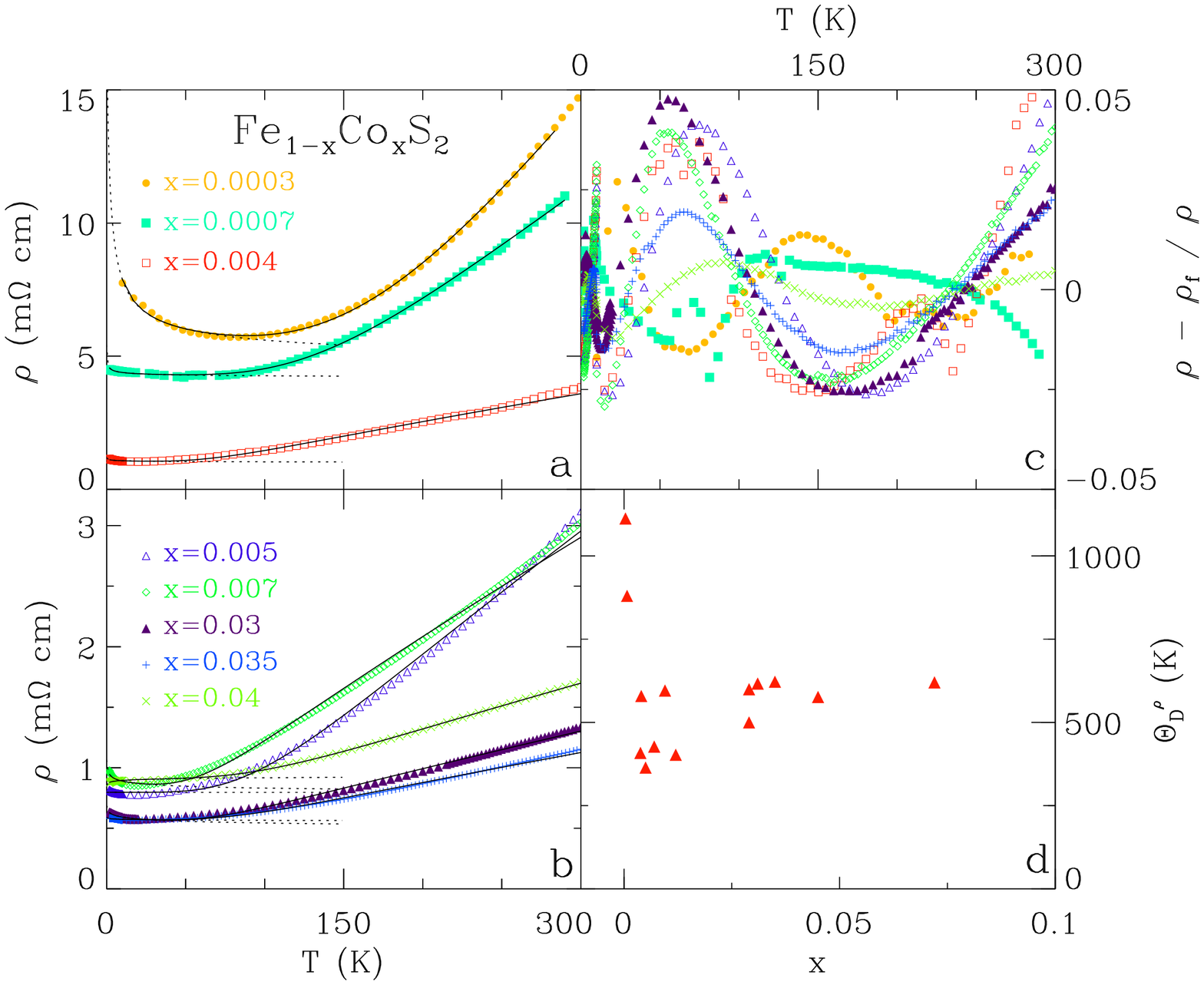}%
  \caption{\label{fig:rhoplv3} (Color online) Temperature dependence of
the resistivity; Bloch model. a) The resistivity, $\rho$ for 3 crystals with Co
concentrations, $x$ indicated in the figure. Solid lines are fits of
the Bloch model of lattice resistance\protect{\cite{zimon,gantmakher}}
to our data. The model includes a residual resistivity, $\rho_0$ added
to a Kondo anomaly determined from fits to the low temperature
resistivity ($T<10 K$) for example see Fig.\
{\protect{\ref{fig:mrpl}}} and displayed below 150 K by the dotted
lines in the figure. b) $\rho$ for 5 representative crystals with
$x$'s indicated in the figure. Dashed and dotted lines are the same as
in frame a. c) $\rho - \rho_f / \rho$, deviation of the resistivity
from the fits, $\rho_f$.  d) Transport Debye temperature,
$\Theta_D^{\rho}$ as determined from fits of the model to the data in
frame a and b.  }
\end{figure}

The resistivity of a few representative samples is shown in
Fig.~\ref{fig:rhoplv3} (a) and (b) where the temperature dependence of
the resistivity appears to be almost linear over a wide range of
temperatures. In the usual description of metallic transport, based in
part on Matthiessen's rule, carrier scattering due to impurities is
considered temperature independent, while the scattering from the
thermally activated phonons is highly $T$-dependent. The
standard\cite{zimon,gantmakher}, semiclassical, Debye treatment of
carrier-phonon scattering predicts a $\rho\propto
(T/\Theta_D^{\rho})^5$ behavior for $T<\Theta_D^{\rho}$ evolving into
a $\rho\propto T/\Theta_D^{\rho}$ form for $T>\Theta_D^{\rho}$. Here
$\Theta_D^{\rho}$ is the transport Debye temperature. The Debye
temperature for iron pyrite is 610 K\cite{robie} and we estimate the
transport Debye temperature, $\Theta_D^{\rho} = 2\hbar k_F s/k_B$,
where $s$ is the speed of sound (8980 m/s), to be between 200 and 250
K\cite{gantmakher,kasami}. To test if the resistivity we measure in
Fe$_{1-x}$Co$_x$S$_2$ can be described as being due to carrier
scattering from phonons, we have fit the full Debye form to our
data. We have included a Kondo form, Eq.~\ref{eq:hamann}, consistent
with the the low-$T$ data, to our model in order to accurately model
the resistivity at $T< 20$ K. In this procedure we have varied the
impurity resistivity, $\rho_0$, the carrier mass enhancement, and
$\Theta_D^{\rho}$, while keeping the carrier density fixed to the
value determined from the Hall effect measurements and assuming a
spherical Fermi surface to estimate $k_F$, in order to find the best
representation of our data. The results of this procedure are shown in
Fig.~\ref{fig:rhoplv3} a and b by the solid lines. The difference
between the data and model is presented as a percentage of the
measured resistivity in frame c of the figure, and the best fit values
of $\Theta_D^{\rho}$ displayed in frame d. Best fit values of the
carrier mass ranged between 1 and 1.5 times the bare electron
mass. From the quality of the fits shown in the figure, we conclude
that the semiclassical treatment adequately describes the resistivity
of our samples over a wide range of $T$ with a few notable exceptions
and qualifications. The first, as is apparent in frame c of the
figure, is that the model does not accurately describe the behavior
below 100 K.  In particular, for samples with $x$ close to $x_c$, the
model deviates substantially from the data near 50 K. In addition, the
values of $\Theta_D^{\rho}$ determined from our fitting procedure for
crystals with $x<x_c$ are much larger than the Debye temperature of
pure FeS$_2$ which would be surprising if true.

\begin{figure}[htb]
  \includegraphics[angle=90,width=3.2in,bb=60 85 550
  685,clip]{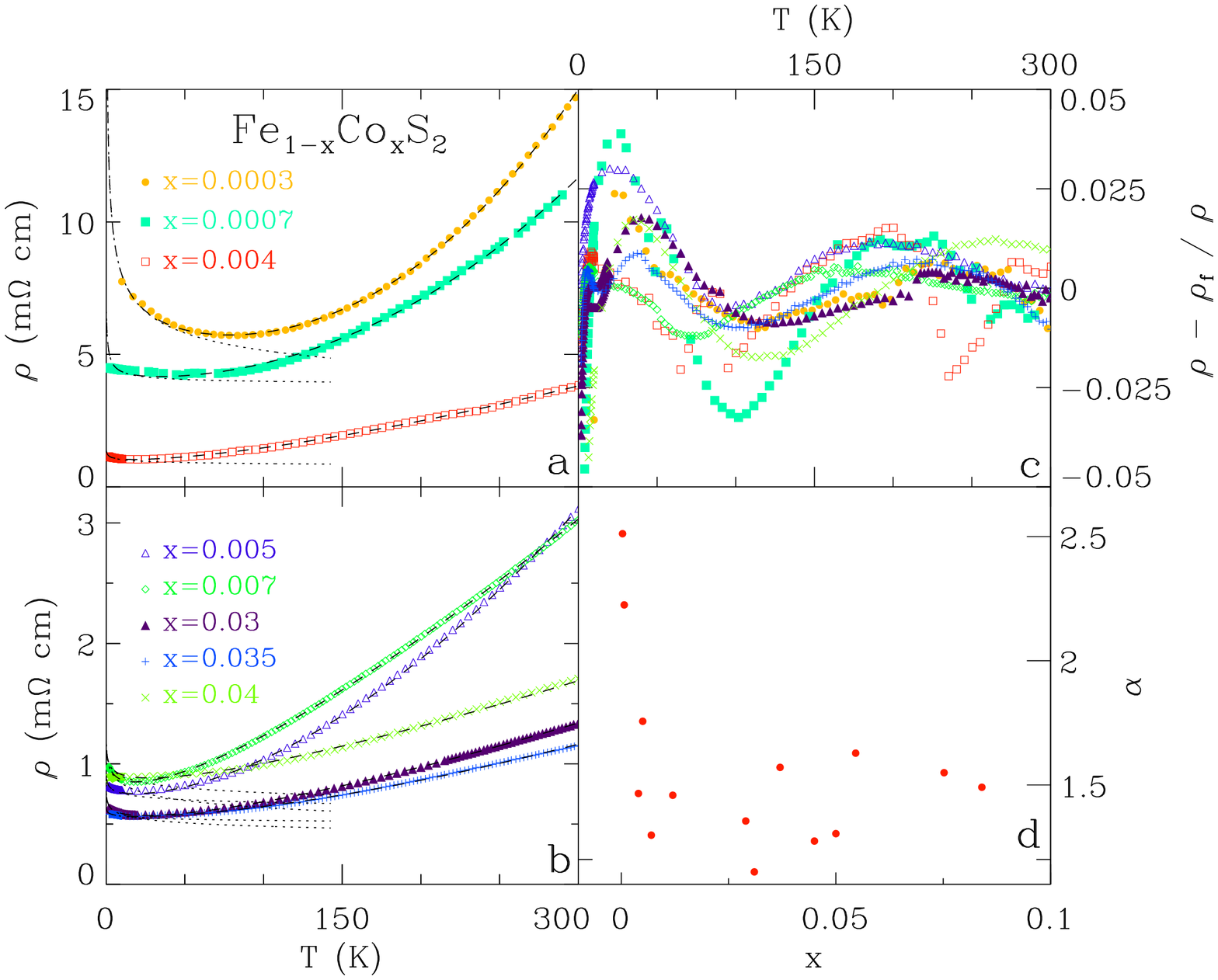}%
  \caption{\label{fig:rhopl} (Color online) Temperature dependence of
the resistivity; power-law model a) The resistivity, $\rho$ for 3
crystals with Co concentrations, $x$ indicated in the figure. Dashed
lines are fits of a model that includes a residual resistivity,
$\rho_0$ added to a Kondo anomaly and a simple power law form,
$AT^{\alpha}$. The Kondo form was determined from fits to the low
temperature resistivity ($T<10 K$) for example see Fig.\
{\protect{\ref{fig:mrpl}}} and displayed below 150 K by the dotted
lines in the figure. b) $\rho$ for 5 representative crystals with
$x$'s indicated in the figure. Dashed and dotted lines are the same as
in frame a. c) $\rho - \rho_f / \rho$, deviation of the resistivity
from the fits, $\rho_f$, showing that for crystals with $x$ in
proximity to $x_c$ a power law form with $\alpha = 1.6 \pm 0.1$
describes the data well over an extended temperature range. d)
$\alpha$ as determined from fits of the model to the data in frame a
and b.}
\end{figure}

We have considered a simple alternative to the standard phonon
scattering model of resistivity motivated by our observation that the
resistivity data appear to have a nearly linear $T$-dependence over a
wide range in $T$.  This alternative, demonstrated in
Fig.~\ref{fig:rhopl}, consists of a simple power-law form, $\rho =
\rho_0 + A T^{\alpha}$, added to the same Kondo description of the
low-$T$ data (Eq.~\ref{eq:hamann}) as in Fig.~\ref{fig:rhoplv3}. The
results of this procedure where $\alpha$, $A$, and $\rho_0$ are
allowed to vary is represented in frames a and b of
Fig.~\ref{fig:rhopl} by the dashed lines. The ability of such a simple
model to reproduce the data is remarkable. In fact, the difference
between the data and model shown in frame c of Fig.~\ref{fig:rhopl} is
smaller for samples with $x\sim x_c$ than the much more complex Debye
model of Fig.~\ref{fig:rhoplv3}, although some systematic differences
remain below 100 K. The best fit values of the temperature exponent,
$\alpha$, are shown in frame d where $\alpha$ is seen to change from
about 2.5 for small $x$ ($x=3\times 10^{-4}$) to values between 1.2
and 1.6.

Why such a simple expression does such an accurate job of describing
our resistivity data is not clear. However, we suggest that the
formation of a magnetically ordered state at low-$T$ implies that
scattering from magnetic fluctuations plays an important role in
determining $\rho(T)$. The large reduction of the residual resistivity
by the application of moderately sized magnetic fields, a decrease of
45\% of $\rho_0$ with a 9 T field as seen in Fig.~\ref{fig:mrfspl},
supports the idea that magnetic scattering is a substantial fraction
of $\rho_0$. Our observation that the RRR$\approx 4$ for samples near
$x_c$, along with the observation that the Debye model for carrier
scattering from phonons has systematic differences with the data, and
the quality of fits of a simple power law temperature dependence to
$\rho(T)$, all indicate magnetic fluctuations to be an important
contribution to the carrier scattering in this range of $x$ over a
wide temperature range.

\begin{figure}[htb]
  \includegraphics[angle=90,width=3.0in,bb=85 100 510
  700,clip]{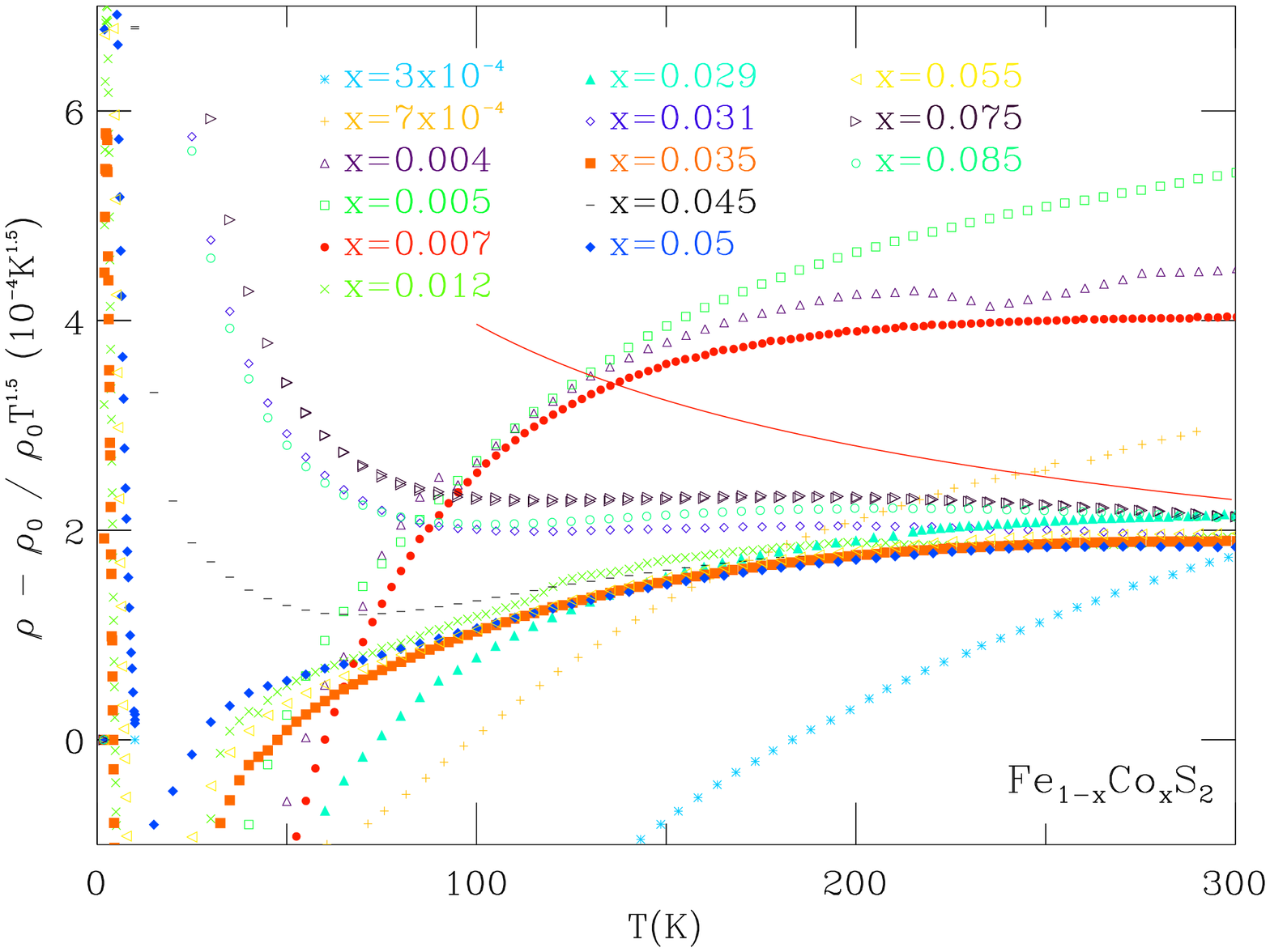}%
  \caption{\label{fig:rhosys2} (Color online) Reduced resistivity. a)
The resistivity, $\rho$, after subtraction of the resistivity at 4 K,
$\rho_0$, normalized by $\rho_0$ and divided by $T^{1.5}$ vs.\
temperature, $T$, for 14 representative crystals with $x$'s identified
in the figure. All samples except those with $x$ nearest to $x_c$ have
values that approach $2\times 10^{-4}$ K$^{-1.5}$ while those with
$x\sim x_c$ display values larger than $4\times 10^{-4}$
K$^{-1/5}$. The red line in the figure represents a linear temperature
dependence, such as that expected for phonon scattering at
$T$ larger then the Debye temperature.}
\end{figure}

The simple power-law analysis of $\rho(T)$ at $T> 20$ K has
demonstrated that the resistivity of our samples closely resembles a
$\rho=\rho_0 +AT^{\alpha}$ dependence with $\alpha \approx 1.5\pm 0.1$
in a wide range of $x$ and $T$. In order to assess the systematics and
validity of this model we have plotted the quantity $(\rho - \rho_0) /
(\rho_0 * T^{1.5})$ in Fig. \ref{fig:rhosys2}. In this way we can
compare scattering rates in samples with very different carrier
concentrations, and remove the error associated with the measurement
of the geometry of our crystals. If the power-law expression were to
strictly hold, this quantity would yield the parameter $A$ divided by
the residual resistivity. Dividing by $\rho_0$ normalizes $A$ by
eliminating the simple effects of carrier concentration and impurity
scattering rate changes with $x$. Thus, the figure isolates the
magnitude of this quantity related to the $T$-dependent part of the
scattering rate of the carriers. The figure demonstrates that this
value is very closely clustered about $(\rho - \rho_0) / (\rho_0
T^{1.5})=2\times 10^{-4}$ K$^{-1.5}$ for nearly all of our
samples. The obvious exceptions are the crystals with $x$ nearest to
$x_c$ on the paramagnetic side where a value of $(\rho - \rho_0) /
(\rho_0T^{1.5}) = 4\times 10^{-4}$ to 6$\times 10^{-4}$ K$^{-1.5}$ is
found, reflecting the observation of a much larger RRR in these
samples.

It appears from this simple analysis that the $T$-dependent scattering
rate above $\sim 150$ K is extraordinarily independent of $x$ for
$x>x_c$ while is at least twice as large for $x\simeq x_c$. From this
we conclude that there is a carrier scattering mechanism that is
enhanced very near the critical concentration for magnetism in this
system. Given the power-law like dependence on $T$, we assert that
carrier scattering from magnetic fluctuations contributes a
significant fraction of the resistivity over the entire temperature
range we measure.

Resistivities, and thus by implication charge carrier scattering
rates, that display similar temperature dependent power law behaviors
for $T<20$ K have been documented for nearly ferromagnetic and weakly
ferromagnetic metals such as Pd$_{1-x}$Ni$_x$\cite{nicklas} and
ZrZn$_2$\cite{smith}, where scattering from spin fluctuations produces
a $T^{5/3}$ dependence of $\rho$. This behavior is observed in highly
itinerant magnets where a Stoner-Wholfarth model of
magnetism\cite{moriya} is appropriate and is thought to signify the
emergence of a marginal Fermi liquid ground state\cite{varma}. A
second group of materials well known for having power-law
$T$-dependent resistivities, $\rho=\rho_0 + AT^{\alpha}$, with small
$\alpha$, for $T<10 K$, are those near quantum critical points (QCP),
such as a zero-$T$ magnetic critical point. Examples include MnSi
under pressure\cite{mathur,doiron},
Ce(Cu$_{1-x}$Au$_x$)$_6$\cite{schroder}, and
YbRh$_2$Si$_2$\cite{custers}. None of these examples display power-law
behavior above about 20 K. One exception is the high temperature
superconducting oxides of copper where $\rho \propto T^{1.5}$ has been
documented for over-doped samples of La$_{2-x}$Sr$_x$CuO$_4$ over an
enormous temperature range\cite{takagi}. How the temperature dependent
resistivity of Fe$_{1-x}$Co$_x$S$_2$ we observe here fits into this
evolving story is not clear. However, we point out that all of these
materials reside near a magnetic phase transition that has been tuned
toward zero temperature.

A simple picture that can explain the $T^{1.5}$ temperature dependence
is to assume that the electrons are scattered by spin wave excitations
of the disordered system.  These spin waves are the excitations around
the Griffiths phase ordered state.  We assume that the density of
states of spin waves in the disordered system varies as $\sqrt{E}$ as
it does in a clean system, despite the fact that the spin waves do not
have a well defined momentum.  In calculating the scattering of an
electron we include all spin wave energies up to about $k_BT$, leading
to a net scattering rate that varies as $T^{1.5}$.  Because the spin
excitations have no well defined momentum, even low energy scattering
events can result in large momentum changes for the electron, so the
transport lifetime is similar to the single particle lifetime, unlike
the case in clean materials where the phonon scattering varies as $T^5$
instead of $T^3$ because of the momeuntum conservation in the
scattering. This scattering process will persist well above the Curie
temperature, as it does in clean materials where neutron scattering
typically sees a well defined spin wave peak even for temperatures
well above the onset of order.

\section{Discussion and Conclusions \label{sec:discuss}}

In Fe$_{1-x}$Co$_x$S$_2$ we have previously observed that the Co
dopants each contribute localized magnetic moments to the slightly
diamagnetic insulator FeS$_2$\cite{guo,guoprb1,jarrett}. These moments
tend to form magnetic clusters and at low $T$ we have observed the
emergence of a disordered ferromagnetic phase at $x_c=0.007\pm
0.002$. Our Hall effect data presented here reveal a density of
electrical carriers of only 10 to 30\% of the Co concentration
indicating a substantial localization of electrons donated by the Co
substitution. A slight curvature evident in the plots of the Hall
potential vs.\ the magnetic field may indicate the existence of a
small population of compensating holes. The reasons for the low
carrier densities is not clear at this time but may reflect a
significant substitutional-related disorder. In the magnetically
ordered samples we observe an anomalous Hall effect with a very large
coefficient ($R_S$) which decreases with $x$.

Measurements of the low-$T$ conductivity show that the charge carriers
interact substantially with the magnetic moments in this system. We
have presented evidence that the Kondo effect dominates the carrier
transport for $x\le x_c$ for $T< 10$ K and, as is the case for both
simple metals with magnetic impurities and Kondo lattice
systems\cite{hamann,rizzuto,fisk}, the resistivity saturates at low
$T$. We point out that the carrier densities estimated from the Hall
effect are sufficient to screen only a small fraction of the doping
induced magnetic moments, leaving the system highly underscreened or
undercompensated. The temperature independent $\rho$ at the lowest
$T$'s indicate a Fermi liquid ground state with contributions from
quantum interference effects commonly observed in doped semiconductors
surprisingly absent.

The usual indicator for the observation of quantum contributions to
the conductivity, $k_F\ell$ is well within the range where such
effects are expected in our data. The reason for the absence of such
contributions is not completely clear. We speculate that the charge
carrier inelastic scattering rate does not become much smaller than
the elastic scattering rate at low temperatures as is required for the
observation of quantum effects. The source of the inelastic scattering
most likely involves the magnetic degrees of freedom associated with
the nearly magnetically ordered (for $x<x_c$), or the spin-glass-like
or disordered ferromagnetic, phase (for $x>x_c$). As we have asserted
previously, the disorder associated with the chemical substitution
leads to the formation of a magnetic Griffiths phase which may provide
an enormous density of low energy excitations and a source for
inelastic scattering for the charge carriers. The application of a
magnetic field substantially decreases the resistivity of our
Fe$_{1-x}$Co$_x$S$_2$ crystals at
low-$T$\cite{schlottmann,andraka}. This is typical of magnetic
materials since the field gaps the low frequency magnetic fluctuation
spectrum and removes much of the inelastic carrier scattering. Despite
the decreased scattering with the application of magnetic fields, we
still do not observe any quantum interference effects.

The low temperature transport properties of Fe$_{1-x}$Co$_x$S$_2$ we
outline above are substantially different from those of Co and Mn
substituted FeSi, a second magnetic semiconducting system where
detailed low-$T$ measurements have been carried
out\cite{manyala,manyala2,manyala3}. While both of the nominally pure
compounds, FeS$_2$ and FeSi, have small band gaps and are nonmagnetic
having very small magnetic susceptibilities, both become magnetic upon
electron doping via Co substitution. However, electron doping of FeSi
results in a highly itinerant helimagnetic ground state where standard
quantum contributions to the conductivity are dominant. Clearly this
is distinct from what we measure here in Fe$_{1-x}$Co$_x$S$_2$. Hole
doping FeSi by way of Mn substitution is similar to
Fe$_{1-x}$Co$_x$S$_2$ in that it was found to display
undercompensation\cite{manyala3}. Here, the Mn dopants contribute a
spin-1 moment and a single spin-1/2 hole carrier. The resistivity of
these compounds remained $T$-dependent down to the lowest
temperatures, however, the temperature dependence was not the
$T^{1/2}$ behavior found in standard semiconducting systems such as
Si:P\cite{rosenbaum,sarachik}. The NFL behavior discovered in
Fe$_{1-x}$Mn$_x$Si is in agreement with calculations predicting a
singular Fermi liquid in the undercompensated Kondo
model\cite{coleman,coleman2,coleman3,schlottmann2}. In addition, the
application of modest magnetic fields reduces the inelastic scattering
and restores the typical temperature and field dependencies expected
for quantum interference effects\cite{manyala}. We remain puzzled as
to the cause of dramatic differences in the transport properties of
these, naively similar, magnetic semiconducting systems.

Finally, the temperature dependence of the resistivity in the
$T$-range where phonon scattering dominates the $\rho$ of typical
metals shows some peculiarities. In particular, the temperature
dependence of the resistivity is enhanced for samples with $x$ very
near $x_c$, having a RRR twice as large as samples on either side of
$x_c$. Although a semiclassical model of carrier-phonon scattering
describes the broad features of $\rho(T)$ our models show systematic
differences with the data below 100 K for the samples with $x\sim
x_c$. We have demonstrated that a simple power-law form surprisingly
describes the data at least as well as the semiclassical
model\cite{zimon,gantmakher} in this range of $x$.  We speculate that
the resistivity of these compounds results, in part, from magnetic
fluctuation scattering of the carriers, even at temperatures
approaching room $T$.

In summary, Co doping of FeS$_2$ adds both local magnetic moments and
electron charge carriers to this band gap insulator. Our data indicate
that although the magnetic and thermodynamic properties of
Fe$_{1-x}$Co$_x$S$_2$ for $0 \le x \le 0.085$ are dominated by the
presence of Griffiths phases at low temperatures, the transport is
Fermi liquid like. It appears that there is a Kondo coupling of the
carriers and local magnetic moments that produces a temperature
independent resistivity below $\sim 0.5$ K. Despite the small
mean-free-path of the doped charge carriers we observe no indication
of quantum corrections to the conductivity which we speculate is due
to a large inelastic scattering rate of the carriers. Thus, our data
indicate that there may be a range over which the inelastic scattering
rate is too large for quantum contributions to be observable, and yet
too small to induce non-Fermi liquid behavior in the transport of
disordered conductors.

We thank I. Vekhter and C. Capan for discussions.  JFD, DPY, and JYC
acknowledge support of the NSF under DMR084376, DMR0449022, and
DMR0756281.


\end{document}